\documentclass[12pt,reqno]{amsart}

\usepackage{fullpage}
\usepackage{graphicx}
\usepackage{amssymb, amsmath, amsxtra}

\newtheorem{thm}{Theorem}[section]
\theoremstyle{plain}
\newtheorem{cor}{Corollary}[section]
\theoremstyle{plain} 
\newtheorem{lem}{Lemma}[section]
\theoremstyle{plain}

\theoremstyle{plain}
\newtheorem{rem}{Remark}[section]
\newtheorem{prop}{Proposition}[section]

\def\X{{\rm X}}
\def\Y{{\rm Y}}
\def\pr{{\rm pr}}

\def\parder#1{\partial_{#1}}
\def\wtil{\widetilde}
\def\Hop{{\mathcal H}}
\def\Eop{{\mathcal E}}
\def\Rop{{\mathcal R}}

\begin{document}
\allowdisplaybreaks[3]

\title{Symmetries and conservation laws of the generalized Krichever-Novikov equation}
\author{
S.C. Anco${}^1$, 
E.D. Avdonina${}^2$, 
A. Gainetdinova${}^2$,
L.R. Galiakberova${}^2$,\\
N.H. Ibragimov${}^{2,3}$, 
T. Wolf${}^1$\\
\\
$^1$Department of Mathematics and Statistics\\ 
Brock University\\
St. Catharines, Ontario, Canada, L2S 3A1.\\
$^2$Laboratory ``Group analysis of mathematical models in natural and engineering sciences"\\
Ufa State Aviation Technical University\\
450000 Ufa, Russia.\\
$^3$Department of Mathematics and Science\\
Blekinge Institute of Technology\\
SE-371 79 Karlskrona, Sweden.
}

\begin{abstract}
A computational classification of contact symmetries 
and higher-order local symmetries that do not commute with $t,x$, 
as well as local conserved densities that are not invariant under $t,x$ 
is carried out for a generalized version of the Krichever-Novikov equation. 
Several new results are obtained. 
First, the Krichever-Novikov equation is explicitly shown to have 
a local conserved density that contains $t,x$. 
Second, apart from the dilational point symmetries known for special cases of
the Krichever-Novikov equation and its generalized version, 
no other local symmetries with low differential order are found to contain $t,x$.
Third, the basic Hamiltonian structure of the Krichever-Novikov equation 
is used to map the local conserved density containing $t,x$ 
into a nonlocal symmetry that contains $t,x$. 
Fourth, a recursion operator is applied to this nonlocal symmetry
to produce a hierarchy of nonlocal symmetries that have explicit dependence on $t,x$. 
When the inverse of the Hamiltonian map is applied to this hierarchy, 
only trivial conserved densities are obtained. 
\end{abstract}

\maketitle

\section{Introduction}

The Krichever-Novikov (KN) equation was introduced in 1979 in the form \cite{KriNov79}
\begin{equation*}
c_t=\frac{3}{8}\frac{1-c_{xx}^2}{c_x}- \frac{1}{2} Q(c) c_{x}^3 + \frac{1}{2} c_{xxx}
\end{equation*}
where $Q(c)$ is expressed in terms of the Weierstrass elliptic function
(see also Ref.\cite{KriNov80}). 
An alternative form of this equation is given by \cite{SviSok82,Nov83,Ibr83kn}
\begin{equation}\label{KNeq}
u_t = u_{xxx} - \frac{3}{2}\,\frac{u_{xx}^2}{u_x} + \frac{p(u)}{u_x}
\end{equation}
where $p(u)$ is an arbitrary quartic polynomial
\begin{equation}\label{quarticKN}
p(u) = C_1 u^4+C_2 u^3+C_3 u^2+C_4 u +C_5
\end{equation}
with constant coefficients.

A main property of the KN equation, 
as shown in Refs.\cite{SviSok82,SviSok82a,IgoMar02},
is that it belongs to the class of integrable evolutionary equations 
in the sense of having an infinite number of higher-order local symmetries 
and higher-order local conserved densities,
which are connected to Hamiltonian structures and recursion operators 
for the equation. 

Among all of these equations, 
the KN equation is singled out by several features. 
For example, 
it is the only nonlinear integrable equation of 
the third order evolutionary form 
$u_t = u_{xxx} + F(u, u_x, u_{xx})$ 
that cannot be mapped to the Korteveg-de Vries (KdV) equation $v_t=v_{xxx}+vv_x$
by a finite-order differential substitution $v=\Phi(u, u_x, u_{xx}, \ldots)$. 
This result follows from the classifications of nonlinear integrable equations
presented in Refs.\cite{IbrSha80a,IbrSha80b,SviSok82,SviSok82a} 
and has been extended to a larger class of equations
$u_t = a(u) u_{xxx} + F(u, u_x, u_{xx})$ in Ref.\cite{Ibr83kn}.
Also, it is the only nonlinear integrable equation in these classes 
that does not have a scaling symmetry. 
More generally, 
the KN equation (for an arbitrary quartic $p(u)$) 
has no known local symmetries that explicitly contain the variables $t,x$,
and likewise none of its local conserved densities that are known to-date \cite{SviSok82a,SviSok82,DemSok08} have explicit dependence on $t,x$. 

This paper is motivated by the question of whether 
the Krichever-Novikov equation admits 
any local symmetries or local conserved densities that have some
essential dependence on $t,x$. 
We will in fact settle this question for a generalization of 
the Krichever-Novikov equation (called the {\em gKN equation})
given by 
\begin{equation}\label{gKN}
u_t = u_{xxx} - \frac{3}{2}\frac{u_{xx}^2}{u_x} + \frac{f(u)}{u_x}
\end{equation}
where $f(u)\not\equiv 0$ is a general function of $u$. 
This equation \eqref{gKN} is integrable only when $f(u)$ is 
a quartic polynomial \eqref{quarticKN},
coinciding with the KN equation. 
The point symmetries of the gKN equation have been recently classified 
in Ref.\cite{LevWinYam11,BruGan11}. 

Symmetries that explicitly contain $t,x$ are related to 
master symmetries \cite{Olv}, 
while conserved densities that explicitly contain $t,x$ are typically
important in the study of solitons and other exact solutions. 

For the gKN equation \eqref{gKN}, 
we will classify its admitted contact symmetries 
and higher-order local symmetries up to differential order six 
that do not commute with translations on $t,x$,
as well as all of its admitted local conserved densities 
up to differential order three. 
Our results yield new conserved densities including one containing $t,x$ 
that holds when the function $f(u)$ is an arbitrary quartic polynomial. 
Thus, we establish that the KN equation \eqref{KNeq}--\eqref{quarticKN} 
does possess a local conserved density with essential dependence on $t,x$. 
We also rule out the existence of any local symmetries 
(with low differential order) that explicitly contain $t,x$, 
other than the known dilational point symmetries that hold only 
for certain (scaling homogeneous) functions $f(u)$. 

Our symmetry and conserved density classifications
are derived in section~\ref{classifysymms} and section~\ref{classifyconslaws},
respectively. 
In section~\ref{symmaction}, 
we work out the action of the point symmetries on the conserved densities. 

The main results appear in section~\ref{nonlocal}. 
We first use the new local conserved density containing $t,x$ to produce 
a nonlocal symmetry from the basic Hamiltonian structure of the KN equation. 
Next we use a recursion operator of the KN equation to generate
a hierarchy of nonlocal symmetries having explicit dependence on $t,x$. 
No such symmetries have been previously found for the KN equation. 
We also discuss the Hamiltonian structure of this hierarchy 
and show that, surprisingly, none of the higher-order nonlocal symmetries 
arise from conserved densities through the basic Hamiltonian structure of the KN equation. 

We make some concluding remarks in section~\ref{remarks}.

From a computational viewpoint, 
the problem of classifying symmetries and conserved densities consists of 
solving a linear system of determining equations 
whose unknowns are functions of $t$, $x$, $u$, and derivatives of $u$
(up to some finite differential order), together with the function $f(u)$. 
Most of the computation involves solving equations that are linear in the unknowns. 
After all dependencies of the unknowns 
on $t,x$ and derivatives of $u$ are determined from these equations, 
the remaining equations involve only the dependencies on $u$,
including the function $f(u)$. 
The solution of these final equations is inherently a nonlinear problem, 
but the equations can be factorized such that the function $f(u)$ 
can be determined by solving an ODE (possibly nonlinear)
while the dependencies of the unknowns on $u$ can be determined by solving 
some linear equations. 
We carry out the computations by using the computer algebra programs
{\sc LiePDE} \cite{Wol93,Wol02a} to compute symmetries 
and {\sc ConLaw} \cite{Wol02a,Wol02b} to compute conservation laws.
We have also verified the results by doing an interactive computation
in Maple which also gave a more compact form for the solutions 
with fewer redundant special cases.

\section{Classification of symmetries}
\label{classifysymms}

To begin, consider the Lie symmetry group of the gKN equation \eqref{gKN}. 
Since the gKN equation involves only a single dependent variable $u$, 
its Lie symmetry group comprises point symmetries and contact symmetries
\cite{Ibr83,Olv,1stbook,2ndbook}. 

A {\em point symmetry} of the gKN equation \eqref{gKN} is 
a group of transformations on $(t,x,u)$ 
given by an infinitesimal generator
\begin{equation}\label{pointsymm}
\X=\tau(t,x,u)\parder{t} +\xi(t,x,u)\parder{x} +\eta(t,x,u)\parder{u}
\end{equation}
whose prolongation satisfies 
\begin{equation}
\pr\X \Big(u_t - u_{xxx} +\frac{3}{2}\frac{u_{xx}^2}{u_x} -\frac{f(u)}{u_x}\Big)
=0 
\end{equation}
for all solutions $u(t,x)$ of the gKN equation \eqref{gKN}. 
When acting on solutions, 
any point symmetry \eqref{pointsymm} is equivalent to 
an infinitesimal generator with the {\em characteristic form} 
\begin{equation}\label{pointsymmchar}
\hat\X=P\parder{u}, 
\quad 
P =\eta(t,x,u)-\tau(t,x,u) u_t-\xi(t,x,u) u_x
\end{equation}
where the characteristic functions $\eta$, $\tau$, $\xi$ are determined by 
\begin{equation}\label{symmdeteq}
\begin{aligned}
0 & 
= \pr\hat\X \Big(u_t - u_{xxx} +\frac{3}{2}\frac{u_{xx}^2}{u_x} -\frac{f(u)}{u_x}\Big)
\\& 
= D_t P-D_x^3 P +3 \frac{u_{xx}}{u_x}D_x^2 P-\frac{3}{2}\frac{u_{xx}^2}{u_x^2} D_x P +\frac{f(u)}{u_x^2} D_x P -\frac{f'(u)}{u_x} P 
\end{aligned}
\end{equation}
holding for all solutions $u(t,x)$ of the gKN equation \eqref{gKN}. 
This formulation is useful for doing computations 
and for considering extensions to contact symmetries and higher-order symmetries,
as well as for making a connection with conserved densities. 

A {\em contact symmetry} extends the definition of invariance \eqref{symmdeteq}
by allowing the transformations to depend 
essentially on first order derivatives of $u$,
as given by an infinitesimal generator with characteristic form 
\begin{equation}\label{contactsymmchar}
\hat\X=P(t,x,u,u_t,u_x)\parder{u} . 
\end{equation}
The corresponding transformations on $(t,x,u,u_t,u_x)$ are given by 
\begin{equation}\label{contactsymm}
\X=\tau\parder{t} +\xi\parder{x} +\eta\parder{u} +\eta^t\parder{u_t} +\eta^x\parder{u_x}
\end{equation}
where
\begin{equation}\label{contactsymmform}
\tau= -P_{u_t},
\quad
\xi= -P_{u_x},
\quad
\eta= P-u_t P_{u_t} -u_x P_{u_x},
\quad
\eta^t= P_{t} + u_t P_u,
\quad
\eta^x= P_{x} + u_x P_u, 
\end{equation}
which follows from preservation of the contact condition $du =u_t dt + u_x dx$.
Note that a contact symmetry reduces to a (prolonged) point symmetry 
if and only if $P$ is a linear function of $u_t$ and $u_x$. 

The set of all infinitesimal point and contact symmetries 
admitted by the gKN equation \eqref{gKN} 
inherits the structure of a Lie algebra under commutation of the operators $\X$.
For a given (sub)algebra of point or contact symmetries,
the corresponding group of transformations has a natural action \cite{Olv,1stbook,2ndbook}
on the set of all solutions $u(t,x)$. 

To classify all of the contact symmetries (and point symmetries) 
admitted by the gKN equation \eqref{gKN},
we first substitute a general characteristic function $P(t,x,u,u_t,u_x)$ 
into the symmetry determining equation \eqref{symmdeteq}. 
Next we eliminate $u_{xxx}$, $u_{txxx}$, $u_{xxxx}$ 
through writing the gKN equation in the solved form 
\begin{equation}\label{u3x}
u_{xxx} = u_t - \dfrac{3}{2}\dfrac{u_{xx}^2}{u_x} + \dfrac{f(u)}{u_x}
\end{equation}
and doing the same for its differential consequences. 
The determining equation \eqref{symmdeteq} then splits 
with respect to $u_{xx}$, $u_{tx}$, $u_{tt}$, $u_{txx}$ 
into a linear overdetermined system of equations on $P(t,x,u,u_t,u_x)$. 
We find that this system contains the equations 
\begin{equation}
P_{u_t u_t}=0,
\quad
P_{u_x u_x}=0,
\quad
P_{u_t u_x}=0,
\end{equation}
which imply that $P$ is linear in $u_t$ and $u_x$. 
Hence $P$ reduces to the characteristic form 
for a point symmetry \eqref{pointsymmchar}. 
We then find that the remaining equations in the system are given by 
\begin{gather}
\tau_u=0 , 
\quad
\tau_x=0 ,
\quad
\xi_u=0 , 
\quad
\xi_{xxx}-\xi_t=0 ,
\quad
\tau_t-3\xi_x=0 ,
\label{tauxieqns}\\
\eta_t=0 ,
\quad
\eta_x=0 ,
\quad
\eta_{uuu}=0 ,
\label{etaeqns}\\
2f(u)\left(\eta_u-2\xi_x\right)-f'(u)\eta=0 . 
\label{feqn}
\end{gather}
These equations \eqref{tauxieqns}--\eqref{feqn} are straightforward to solve. 
If $\eta=0$, then we have 
\begin{equation}
\tau=C_1,
\quad
\xi=C_2 . 
\end{equation}
When $\eta\neq 0$, then instead we obtain
\begin{equation}
\tau=3C_1t+C_2,
\quad
\xi=C_1x+C_3,
\quad
\eta=C_4u^2+C_5u+C_6,
\end{equation}
together with the condition 
\begin{equation}\label{fODE}
\frac{f'(u)}{f(u)}=\frac{2(2C_4u+C_5-2C_1)}{C_4u^2+C_5u+C_6} . 
\end{equation}
The integration of ODE \eqref{fODE} splits into five distinct cases:
(i) $C_4=C_5=0$; (ii) $C_4=0$, $C_5\neq0$; 
(iii) $C_4\neq0$, $C_5^2-4C_4C_6=0$;
(iv) $C_4\neq0$, $C_5^2-4C_4C_6>0$;
(v) $C_4\neq0$, $C_5^2-4C_4C_6<0$. 
This leads to the following classification result. 

\begin{thm}\label{class-pointsymm}
(i) For any $f(u)$, the gKN equation \eqref{gKN} admits no contact symmetries.
\newline 
(ii) The point symmetries admitted by the gKN equation \eqref{gKN} 
for arbitrary $f(u)$ consist of 
\begin{equation}\label{pointsymm-general}
\X_1=\parder{t},
\quad
\X_2=\parder{x} . 
\end{equation}
(iii) The gKN equation \eqref{gKN} admits additional point symmetries 
only for the following $f(u)\not\equiv 0$:
\begin{align}
{\rm (a)}\qquad
\label{pointsymm-a}
&\begin{aligned}
& f(u)=C \exp(4au),
\quad
a\neq0
\\ 
& \X_{3\rm a}=-3at\parder{t} -ax\parder{x} +\parder{u} 
\end{aligned}
\\\nonumber\\
{\rm (b)}\qquad
\label{pointsymm-b}
&\begin{aligned}
& f(u)=C (u+b)^{2-4a},
\quad
a\neq -1/2
\\
& \X_{3\rm b}=3at\parder{t}+ax\parder{x} + (u+b)\parder{u} 
\end{aligned}
\\\nonumber\\
{\rm (c)}\qquad
\label{pointsymm-c}
&\begin{aligned}
& f(u)=C (u+b)^4 \exp(4a/(u+b)),
\quad
a\neq 0
\\ 
& \X_{3\rm c}=3at\parder{t}+ax\parder{x} + (u+b)^2\parder{u} 
\end{aligned}
\\\nonumber\\
{\rm (d)}\qquad
\label{pointsymm-d}
&\begin{aligned}
& f(u)=C(c+b+u)^{2+2a/c}(c-b-u)^{2-2a/c},
\quad
c\neq0,
\quad
a\neq\pm c
\\
& \X_{3\rm d}=3at\parder{t}+ax\parder{x} + ((u+b)^2-c^2)\parder{u} 
\end{aligned}
\\\nonumber\\
{\rm (e)}\qquad
\label{pointsymm-e}
&\begin{aligned}
& f(u)=C ((u+b)^2+c^2)^2\exp\left((4a/c)\arctan((u+b)/c)\right),
\quad
c\neq0
\\
& \X_{3\rm e}=-3at\parder{t} -ax\parder{x}+((u+b)^2+c^2)\parder{u} 
\end{aligned}
\\\nonumber\\
{\rm (f)}\qquad
\label{pointsymm-f}
&\begin{aligned}
& f(u)=C (u+b)^4
\\
& \X_{3\rm f}=-(3t/2)\parder{t}-(x/2)\parder{x} + (u+b)\parder{u} 
=\X_{3\rm b}|_{a=-1/2} ,
\\
& \X_{4\rm f}=(u+b)^2\parder{u} 
=\X_{3\rm c}|_{a=0} 
\end{aligned}
\end{align}
\end{thm}

Modulo equivalence transformations, 
this classification can be easily checked to reduce to the classification of
point symmetries of the gKN equation stated in Ref.\cite{LevWinYam11}. 
(The classification presented in Ref.\cite{BruGan11} is missing cases \eqref{pointsymm-c} and \eqref{pointsymm-d}.)
The equivalence transformations of the gKN equation,
other than symmetry transformations, 
consist of a scaling on $t,x$, and $f(u)$, 
as well as a Mobius transformation on $u$ and $f(u)$ \cite{LevWinYam11,DemSok08}. 
Here, we are interested in getting an explicit classification without 
having to change the form of $f(u)$ under equivalence transformations. 

A {\em higher-order symmetry} of the gKN equation \eqref{gKN}
is an infinitesimal generator 
\begin{equation}\label{symmchar}
\hat\X=P(t,x,u,u_x,u_{xx},\ldots)\parder{u}
\end{equation}
whose characteristic function $P$ has a differential order of at least two,
satisfying the symmetry determining equation \eqref{symmdeteq}
for all solutions $u(t,x)$. 
Existence of higher-order symmetries that commute with translations on $t,x$ 
is connected with integrability. 
Since the gKN equation is integrable only when it coincides with 
the KN equation \eqref{KNeq}--\eqref{quarticKN}, 
we expect that such symmetries $\hat\X=P(u,u_x,u_{xx},\ldots)\parder{u}$
will be admitted only when $f(u)$ is a quartic polynomial. 
This still leaves open the possibility for existence of higher-order symmetries
that involve $t,x$ explicitly, which would be related to a master symmetry. 

To look for symmetries that involve $t,x$ explicitly, 
we will now classify all higher-order symmetries \eqref{symmchar}
up differential order six admitted by the gKN equation \eqref{gKN}. 
In particular, 
any such symmetries that are admitted for special $f(u)$ will be determined. 

For computational purposes, 
it will be useful to work with an equivalent representation 
for the symmetries \eqref{symmchar},
in which $u_{xxx},\ldots,u_{xxxxxx}$ are eliminated in $P$
through the solved form of the gKN equation \eqref{u3x}. 
Then we have the following correspondence result. 

\begin{lem}\label{convertsymm}
For the gKN equation \eqref{gKN}, 
symmetries up to differential order six in $x$ derivatives 
\begin{equation}\label{utelimsymm}
\hat\X = P(t,x,u,u_x,u_{xx},u_{xxx},u_{xxxx},u_{xxxxx},u_{xxxxxx})\parder{u}
\end{equation} 
are equivalent to symmetries of the form 
\begin{equation}\label{u3xelimsymm}
\hat\X= \wtil P(t,x,u,u_t,u_x,u_{tx},u_{xx},u_{tt},u_{txx})\parder{u}
\end{equation} 
whose differential order in $t,x$ derivatives is at most three
(where $P_{u_{ttx}}=P_{u_{ttt}}=0$). 
\end{lem}

The determining equation \eqref{symmdeteq} for symmetries \eqref{u3xelimsymm}
splits with respect to $u_{ttx}$, $u_{ttxx}$
into a linear overdetermined system of equations on 
\begin{equation}\label{altsymmchar}
\wtil P(t,x,u,u_t,u_x,u_{tx},u_{xx},u_{tt},u_{txx}) . 
\end{equation} 
This system can solved in a straightforward computational way. 

\begin{thm}\label{class-symmchar}
The gKN equation \eqref{gKN} admits 
a higher-order symmetry characteristic \eqref{altsymmchar}
only when $f(u)\not\equiv 0$ is a quartic polynomial:
\begin{align}
& f(u)=C_1u^4+C_2u^3+C_3u^2+C_4u+C_5
\nonumber\\
& \begin{aligned}
\wtil P = &
u_{txx} -2\frac{u_{tx}u_{xx}}{u_x} +\frac{1}{2}\frac{u_{t}u_{xx}^2}{u_x^2} 
+\frac{1}{2}\frac{u_t^2}{u_x} -\frac{4}{3}\frac{u_{xx}^2 f(u)}{u_x^3} 
+\frac{4}{3}\frac{u_{xx} f'(u)}{u_x} 
\\&\qquad 
-\frac{5}{3}\frac{u_t f(u)}{u_x^2} 
+\frac{8}{9}\frac{f(u)^2}{u_x^3} -\frac{4}{9} u_x f''(u) .
\end{aligned}
\end{align}
\end{thm} 

We next rewrite this symmetry by eliminating $u_t$, $u_{tx}$, $u_{txx}$ 
through the gKN equation \eqref{gKN} and its differential consequences. 
This provides an explicit classification of 
all higher-order symmetries \eqref{utelimsymm} up to differential order six
(in $x$ derivatives). 

\begin{cor}\label{class-symms}
(i) For any $f(u)$, 
the gKN equation \eqref{gKN} admits no symmetries \eqref{utelimsymm} 
of differential order one, two, four, or six. 
\newline
(ii) The gKN equation \eqref{gKN} admits 
a single symmetry \eqref{utelimsymm} of differential order three:
\begin{align}
& 
\hat\X_1 = 
\Big(u_{xxx} -\frac{3}{2}\frac{u_{xx}^2}{u_x} +\frac{f(u)}{u_x}\Big)\parder{u}
= u_t\parder{u}
\label{3rdord-P}
\\
&
f(u) \text{ arbitrary } .
\end{align}
(iii) The gKN equation \eqref{gKN} admits 
a single symmetry \eqref{utelimsymm} of differential order five:
\begin{align}
&
\begin{aligned} 
\hat\X_2 = & 
\Big( u_{xxxxx} -5\frac{u_{xxxx}u_{xx}}{u_x}
-\frac{5}{2}\frac{u_{xxx}^2}{u_x} +\frac{25}{2}\frac{u_{xxx}u_{xx}^2}{u_x^2} 
-\frac{45}{8}\frac{u_{xx}^4}{u_x^3} 
-\frac{5}{3}\frac{u_{xxx} f(u)}{u_x^2} 
\\&\qquad
+\frac{25}{6}\frac{u_{xx}^2 f(u)}{u_x^3} -\frac{5}{3}\frac{u_{xx} f'(u)}{u_x} 
-\frac{5}{18}\frac{f(u)^2}{u_x^3} +\frac{5}{9}u_x f''(u) \Big)\parder{u}
\end{aligned} 
\label{5thord-P}\\
& 
f'''''(u) =0 
\quad 
(f(u)=C_1u^4+C_2u^3+C_3u^2+C_4u+C_5) .
\end{align}
\end{cor} 

This classification establishes that both the KN and gKN equations 
do not possess any higher-order symmetries \eqref{utelimsymm} 
that have at most differential order six (in $x$ derivatives) 
and also involve $t,x$ explicitly.

The symmetry \eqref{5thord-P} of differential order five 
is the first higher-order symmetry in the hierarchy generated by 
the fourth order recursion operator of the KN equation \cite{Sok84},
where the root symmetry for this hierarchy is $\hat\X=u_x\parder{u}$. 
All of the symmetries in the hierarchy commute with translations on $t,x$.

\section{Classification of conservation laws}
\label{classifyconslaws}

A {\em conservation law} of the gKN equation \eqref{gKN}
is a space-time divergence such that
\begin{equation}\label{conslaw}
D_t T(t,x,u,u_t,u_x,\ldots) +D_x X(t,x,u,u_t,u_x,\ldots)=0
\end{equation}
holds for all solutions~$u(t,x)$ of the gKN equation \eqref{gKN}. 
The spatial integral of the conserved density $T$ formally satisfies
\begin{equation}
\frac d{dt} \int_{-\infty}^{\infty} T dx 
= -X\Big|_{-\infty}^{\infty}
\end{equation}
and so if the spatial flux $X$ vanishes at spatial infinity, 
then 
\begin{equation}\label{C}
\mathcal C[u]= \int_{-\infty}^{\infty} T dx=\text{const.}
\end{equation}
yields a conserved quantity for gKN equation \eqref{gKN}.
Conversely, any such conserved quantity arises from
a conservation law \eqref{conslaw}.
Two conservation laws are equivalent if 
their conserved densities $T(t,x,u,u_t,u_x,\ldots)$
differ by a total $x$-derivative $D_x\Theta(t,x,u,u_t,u_x,\ldots)$
on all solutions $u(t,x)$,
thereby giving the same conserved quantity $\mathcal C[u]$ up to boundary terms.
Correspondingly, 
the fluxes $X(t,x,u,u_t,u_x,\ldots)$ of two equivalent conservation laws
differ by a total time derivative $-D_t\Theta(t,x,u,u_t,u_x,\ldots)$ 
on all solutions $u(t,x)$.
A conservation law is called {\em trivial} if 
\begin{equation}\label{trivconslaw}
\begin{aligned}
& T(t,x,u,u_t,u_x,\ldots) = \Phi(t,x,u,u_t,u_x,\ldots) + D_x\Theta(t,x,u,u_t,u_x,\ldots), 
\\
& 
X(t,x,u,u_t,u_x,\ldots) = \Psi(t,x,u,u_t,u_x,\ldots) - D_t\Theta(t,x,u,u_t,u_x,\ldots)
\end{aligned}
\end{equation}
such that $\Phi=\Psi=0$ holds on all solutions $u(t,x)$.
Thus, equivalent conservation laws differ by a trivial conservation law. 

The set of all conservation laws (up to equivalence) 
admitted by the gKN equation \eqref{gKN} 
forms a vector space on which there is a natural action \cite{Olv,BluTemAnc}
by the group of all Lie symmetries of the gKN equation \eqref{gKN}. 

Each conservation law \eqref{conslaw} has 
an equivalent {\em characteristic form} 
in which $u_t$ and all derivatives of $u_t$ are eliminated 
from $T$ and $X$ through use of the gKN equation \eqref{gKN} 
and its differential consequences. 
There are two steps to obtaining the characteristic form. 
First, 
we eliminate $u_t,u_{tx},\ldots$ to get 
\begin{equation}
\hat T = T\big|_{u_t =u_{xxx} - (\frac{3}{2}u_{xx}^2 +f(u))/u_x} 
= T - \Phi,
\quad
\hat X = X\big|_{u_t =u_{xxx} - (\frac{3}{2}u_{xx}^2 +f(u))/u_x} 
= X - \Psi
\end{equation}
so that 
\begin{equation}\label{charconslaw}
\big( D_t\hat T(t,x,u,u_x,u_{xx},\ldots) +D_x\hat X(t,x,u,u_x,u_{xx},\ldots) \big)
\big|_{u_t =u_{xxx} - (\frac{3}{2}u_{xx}^2 +f(u))/u_x} =0
\end{equation}
where 
\begin{align}
& \begin{aligned}
D_t\big|_{u_t =u_{xxx} - (\frac{3}{2}u_{xx}^2 +f(u))/u_x} & = 
\parder{t} 
+ \Big(u_{xxx} - \frac{3}{2}\frac{u_{xx}^2}{u_x} + \frac{f(u)}{u_x}\Big)\parder{u}
\\&\qquad
+ D_x\Big(u_{xxx} - \frac{3}{2}\frac{u_{xx}^2}{u_x} + \frac{f(u)}{u_x}\Big)\parder{u_x} +\cdots
\end{aligned}
\\
& D_x\big|_{u_t =u_{xxx} - (\frac{3}{2}u_{xx}^2 +f(u))/u_x} 
= \parder{x} +u_x\parder{u} +u_{xx}\parder{u_x}+\cdots = D_x 
\end{align}
holds on all solutions of the gKN equation \eqref{gKN}. 
Next, moving off of solutions, 
we use the identity 
\begin{equation}
\begin{aligned}
D_t = & D_t\big|_{u_t =u_{xxx} - (\frac{3}{2}u_{xx}^2 +f(u))/u_x} 
\\&\qquad
+ \Big(u_t -u_{xxx} +\frac{3}{2}\frac{u_{xx}^2}{u_x} -\frac{f(u)}{u_x}\Big)\parder{u}
+ D_x\Big(u_t - u_{xxx} +\frac{3}{2}\frac{u_{xx}^2}{u_x} -\frac{f(u)}{u_x}\Big)\parder{u_x}
+\cdots . 
\end{aligned}
\end{equation}
This yields the characteristic form of the conservation law \eqref{conslaw}
\begin{equation}\label{chareqn}
\begin{aligned}
& D_t\hat T(t,x,u,u_x,u_{xx},\ldots) +D_x\big(\hat X(t,x,u,u_x,u_{xx},\ldots) +\hat \Psi(t,x,u,u_t,u_x,\ldots) \big)
\\
& =Q(t,x,u,u_x,u_{xx},\ldots) \Big(u_t - u_{xxx} +\frac{3}{2}\frac{u_{xx}^2}{u_x} -\frac{f(u)}{u_x}\Big)
\end{aligned}
\end{equation}
holding identically, 
where
\begin{equation}\label{trivX}
\hat\Psi = E_{u_x}(\hat T)\Big(u_t - u_{xxx} +\frac{3}{2}\frac{u_{xx}^2}{u_x} -\frac{f(u)}{u_x}\Big) + E_{u_{xx}}(\hat T)D_x\Big(u_t - u_{xxx} +\frac{3}{2}\frac{u_{xx}^2}{u_x} -\frac{f(u)}{u_x}\Big) + \cdots
\end{equation}
is a trivial flux, 
and where the function 
\begin{equation}\label{TQrelation}
Q=E_u(\hat T) 
\end{equation}
is called a {\em multiplier} (or a {\em characteristic}). 
Here $E_u=\parder{u}-D_x\parder{u_x} +D_x^2\parder{u_{xx}} -\cdots$
denotes the (spatial) Euler operator with respect to $u$. 

From the characteristic equation \eqref{chareqn},
there is a one-to-one relation between 
conserved densities (up to equivalence) 
and multipliers for the gKN equation \eqref{gKN}. 
Note that if a conserved density $\hat T$ has differential order $k\geq 0$,
then the differential order of the corresponding multiplier $Q$ is at most $2k\geq 0$. 
For a conserved density $\hat T$ of minimal differential order $k\geq 0$, 
the corresponding multiplier $Q$ has maximal differential order $2k\geq 0$,
and from the characteristic equation \eqref{chareqn},
the flux $\hat X$ has differential order $k+2\geq 2$. 
We define the {\em differential order of a conservation law} to be 
the smallest differential order among all equivalent conserved densities. 

Multipliers $Q$ are determined by the condition that their product
with the gKN equation is a total space-time divergence. 
Such divergences have the characterization that their variational derivative
with respect to $u$ vanishes identically \cite{Olv,2ndbook}. 
This condition 
\begin{equation}\label{Qvardercond}
\frac{\delta}{\delta u}\bigg( 
\Big(u_t - u_{xxx} +\frac{3}{2}\frac{u_{xx}^2}{u_x} -\frac{f(u)}{u_x}\Big)Q
\bigg)=0
\end{equation}
can be split in an explicit form with respect to $u_t,u_{tx},u_{txx},\ldots$,
which yields the equivalent equations 
\cite{AncBlu97,AncBlu02a,AncBlu02b} 
\begin{equation}\label{adjsymmdeteq}
0 = -D_t Q +D_x^3 Q +3D_x^2 \Big(\frac{u_{xx}}{u_x} Q\Big)
+D_x\Big( \frac{3}{2}\frac{u_{xx}^2}{u_x^2} Q -\frac{f(u)}{u_x^2} Q\Big) -\frac{f'(u)}{u_x} Q
\end{equation}
and
\begin{equation}\label{Helmholtzeq}
Q_u = E_u(Q),
\quad
Q_{u_x} = -E_u^{(1)}(Q),
\quad
Q_{u_{xx}} = E_u^{(2)}(Q),
\quad
\ldots
\end{equation}
holding for all solutions $u(t,x)$ of the gKN equation \eqref{gKN}.
Here $E_u^{(1)} = \parder{u_x}-2D_x\parder{u_{xx}} +3D_x^2\parder{u_{xxx}} -\cdots$
and $E_u^{(2)} = \parder{u_{xx}}-3D_x\parder{u_{xxx}} +6D_x^2\parder{u_{xxxx}} -\cdots$ 
denote higher (spatial) Euler operators \cite{Olv}. 
These equations \eqref{adjsymmdeteq}--\eqref{Helmholtzeq} constitute
the standard determining system for multipliers 
(see also Refs.\cite{Zha,BluCheAnc}). 

The first equation \eqref{adjsymmdeteq} 
is the adjoint of the symmetry determining equation \eqref{symmdeteq},
and its solutions $Q$ are called {\em adjoint-symmetries} 
(or {\em cosymmetries}).
The second equation \eqref{Helmholtzeq} comprises the Helmholtz conditions,
which are necessary and sufficient for $Q$ to be 
an Euler-Lagrange expression \eqref{TQrelation}. 
Consequently, 
multipliers are simply adjoint-symmetries that have a variational form,
and the determination of conservation laws via multipliers is
a kind of adjoint problem \cite{AncBlu97} of the determination of symmetries. 

In this formulation, 
conserved densities and fluxes can be recovered from multipliers 
either by \cite{Wol02b,2ndbook} 
directly integrating the relation \eqref{TQrelation} between $Q$ and $\hat T$, 
or by \cite{Olv,AncBlu02a,AncBlu02b} 
using a homotopy integral formula which expresses $\hat T$ in terms of $Q$
(see also Refs.\cite{DecNiv,PooHer}). 
Alternatively, 
a simple expression for $\hat T$ can be derived from 
the scaling method developed in Ref.\cite{Anc}
applied here to the scaling transformation 
\begin{equation}\label{scaling}
t\rightarrow\lambda^3 t,
\quad
x\rightarrow\lambda x,
\quad
f(u)\rightarrow\lambda^{-4} f(u)
\end{equation}
which belongs to the equivalence group of the gKN equation \eqref{gKN}. 

We will now classify all conservation laws \eqref{charconslaw}
up to differential order three 
admitted by the gKN equation \eqref{gKN}. 
This classification means finding 
all conserved densities $\hat T$ up to differential order three (in $x$ derivatives)
and all fluxes $\hat X$ up to differential order five (in $x$ derivatives), 
or equivalently, 
all multipliers \eqref{TQrelation} up to differential order six
(in $x$ derivatives). 
In particular, 
any such conservation laws that are admitted for special $f(u)$ 
will be determined. 

For computational purposes, 
it will be simpler to use an alternative formulation of this classification,
which arises through expressing the gKN equation in the solved form \eqref{u3x}.
To proceed, 
we first state a useful correspondence result. 

\begin{lem}\label{convertconslaw}
For the gKN equation \eqref{gKN}, 
conservation laws up to differential order three in $x$ derivatives 
\begin{equation}\label{utelimconslaw}
D_t T(t,x,u,u_x,u_{xx},u_{xxx}) + D_x X(t,x,u,u_x,u_{xx},u_{xxx},u_{xxxx},u_{xxxxx})
=0
\end{equation} 
are equivalent to conservation laws 
\begin{equation}\label{u3xelimconslaw}
D_t \wtil T(t,x,u,u_t,u_x,u_{tx},u_{xx})  
+D_x \wtil X(t,x,u,u_t,u_x,u_{tx},u_{xx},u_{tt})
=0
\end{equation} 
whose differential order in $t,x$ derivatives is at most two
(where $\wtil T_{u_{tt}}=0$). 
\end{lem}

The proof involves three main steps. 
First, 
by eliminating $u_{xxx}$, $u_{xxxx}$, $u_{xxxxx}$ through equation \eqref{u3x}, 
we note that the conserved density and the flux 
in a conservation law \eqref{utelimconslaw}
are equivalent to a conserved density $\hat T(t,x,u,u_t,u_x,u_{xx})$
and a flux $\hat X(t,x,u,u_t,u_x,u_{tx},u_{xx},u_{txx})$,
both of which have lower differential order. 
Second, 
we look at the highest $t$-derivative terms 
in the resulting conservation law
\begin{equation}\label{equivconslaw}
D_t \hat T +D_x \hat X =0 . 
\end{equation}
From $D_t\hat T$, we get the term 
$u_{tt}\hat T_{u_{t}}$,
which is second order in $t$ derivatives. 
From $D_x\hat X$, we have
$u_{txxx}\hat X_{u_{txx}}= D_t\Big( u_t - \dfrac{3}{2}\dfrac{u_{xx}^2}{u_x} + \dfrac{f(u)}{u_x} \Big)\hat X_{u_{txx}}$
using equation \eqref{u3x},
which yields the $t$-derivative term 
$u_{tt}\hat X_{u_{txx}}$. 
The conservation law \eqref{equivconslaw} can then hold only if 
the coefficient of $u_{tt}$ vanishes, 
$\hat T_{u_{t}} + \hat X_{u_{txx}} =0$. 
This relation implies 
\begin{equation}\label{Xterms}
\hat X= -u_{txx} \hat T_{u_{t}}+ \hat Y
\end{equation}
for some expression $\hat Y(t,x,u,u_t,u_x,u_{tx},u_{xx})$. 
Finally, 
motivated by the form of the highest derivative term in $\hat X$, 
we subtract a trivial conservation law \eqref{trivconslaw} given by 
\begin{equation}\label{trivterm}
\Theta(t,x,u,u_t,u_x,u_{xx}) = \int \hat T_{u_{t}} du_{xx} .
\end{equation}
This subtraction produces an equivalent conservation law
\begin{equation}\label{u3xelimequiv}
D_t \wtil T +D_x \wtil X =0,
\quad
\wtil T= \hat T -D_x\Theta -\Phi,
\quad
\wtil X= \hat X +D_t\Theta -\Psi
\end{equation} 
where
\begin{equation}
\begin{aligned}
& 
\Phi = 
\Big( u_t -u_{xxx}- \dfrac{3}{2}\dfrac{u_{xx}^2}{u_x} + \dfrac{f(u)}{u_x} \Big)
\hat T _{u_t}
\\
&
D_x\Theta = 
u_{xxx} \hat T_{u_t}
+ \int( \hat T_{xu_t} + u_x \hat T_{uu_t} + u_{xx} \hat T_{u_xu_t} + u_{tx} \hat T_{u_tu_t} ) du_{xx}
\end{aligned}
\end{equation} 
and
\begin{equation}
\begin{aligned}
& 
\Psi = 0
\\
& 
D_t\Theta = 
u_{txx} \hat T _{u_t}
+ \int( \hat T_{tu_t} + u_t \hat T_{uu_t} + u_{tx} \hat T_{u_xu_t} + u_{tt} \hat T_{u_tu_t} )du_{xx} . 
\end{aligned}
\end{equation} 
Hence, on solutions $u(t,x)$ of the gKN equation \eqref{u3x}, 
the equivalent conserved density 
\begin{equation}
\begin{aligned}
\wtil T(t,x,u,u_t,u_x,u_{tx},u_{xx})
= &\hat T -\Big( u_t -\dfrac{3}{2}\dfrac{u_{xx}^2}{u_x} + \dfrac{f(u)}{u_x} \Big)
\hat T_{u_t}
\\&\qquad
- \int( \hat T_{xu_t} + u_x \hat T_{uu_t}+ u_{xx} \hat T_{u_xu_t} + u_{tx}\hat T_{u_tu_t} )du_{xx}
\end{aligned}
\end{equation} 
still has differential order two in $t,x$ derivatives,
while the differential order of the equivalent flux 
\begin{equation}
\hat X(t,x,u,u_t,u_x,u_{tx},u_{xx},u_{tt}) 
= \hat Y 
+ \int( \hat T_{tu_t} + u_t \hat T_{uu_t} + u_{tx} \hat T_{u_xu_t} + u_{tt} \hat T_{u_tu_t} )du_{xx}
\end{equation} 
is lower by one order in $t,x$ derivatives. 
This completes the proof of Lemma~\ref{convertconslaw}. 

It is straightforward to derive that 
a conservation law \eqref{u3xelimconslaw} of mixed differential order two 
has the characteristic form 
\begin{equation}\label{charalteqn}
\begin{aligned}
& D_t \wtil T(t,x,u,u_t,u_x,u_{tx},u_{xx}) 
+D_x \wtil X(t,x,u,u_t,u_x,u_{tx},u_{xx},u_{tt})
\\
& =\wtil Q(t,x,u,u_t,u_x,u_{tx},u_{xx},u_{tt}) \Big(u_t - u_{xxx} +\frac{3}{2}\frac{u_{xx}^2}{u_x} -\frac{f(u)}{u_x}\Big)
\end{aligned}
\end{equation}
holding as an identity,
with the relation 
\begin{equation}\label{Qalt}
\wtil Q=-\wtil X_{u_{xx}} . 
\end{equation}
In this formulation, 
note that the mixed differential order of the multiplier \eqref{Qalt} is two. 

The determining system for these conservation law multipliers \eqref{Qalt} 
is given by the variational derivative condition \eqref{Qvardercond},
which can be split with respect to $u_{xxx}$, $u_{xxxt}$, $u_{xxxx}$, 
$u_{xxxxt}$, $u_{xxxtt}$, $u_{xxxxx}$. 
This splitting yields the adjoint-symmetry equation \eqref{adjsymmdeteq} 
plus additional Helmholtz-type equations given by 
\begin{align}
\hat E_u(\wtil Q) & = 
-\parder{u_{xx}}\Big( \hat D_x\Big(3\dfrac{u_{xx}}{u_x} \wtil Q\Big) + \Big(\dfrac{3}{2}\dfrac{u_{xx}^2}{u_x^2} + \dfrac{f(u)}{u_x^2}\Big) \wtil Q\Big)
+\hat D_x^2 \big(\wtil Q_{u_{xx}}\big)  
\nonumber\\&\qquad
+\hat D_x\Big( \parder{u_{xx}}\Big(\hat D_x \wtil Q - 3\dfrac{u_{xx}}{u_x} \wtil Q\Big) \Big)
+\parder{u_{xx}}\Big( \hat D_x^2 \wtil Q\Big) 
\label{Helmholtzalteq1}\\
-\hat E_u^{(1,t)}(\wtil Q) & = 
2\hat D_x \big(\wtil Q_{u_{tx}}\big) 
+\parder{u_{tx}}\Big( \hat D_x \wtil Q-3\dfrac{u_{xx}}{u_x} \wtil Q\Big)
\label{Helmholtzalteq2}\\
-\hat E_u^{(1,x)}(\wtil Q) & = 
2\hat D_x \big(\wtil Q_{u_{xx}}\big) 
+\parder{u_{xx}}\Big( \hat D_x \wtil Q - 3\dfrac{u_{xx}}{u_x} \wtil Q \Big)
\label{Helmholtzalteq3}\\
\parder{u_{xx}}\Big(\hat E_{u_x}(\wtil Q)\Big) & = 
\parder{u_{xx}}\Big( \hat D_x\big(\wtil Q_{u_{xx}u_{xx}}\big) \Big)
+\parder{u_{xx}}^2\Big(\hat D_x \wtil Q- 3\dfrac{u_{xx}}{u_x} \wtil Q\Big) 
\label{Helmholtzalteq4}
\end{align}
with $\hat D_x$ denoting $D_x$ restricted to solutions 
$u(t,x)$ of the gKN equation \eqref{u3x}. 
Here 
$\hat E_u=\parder{u} -D_t\parder{u_t} -\hat D_x\parder{u_x} 
+D_t^2\parder{u_{tt}} +D_t\hat D_x\parder{u_{tx}} 
+\hat D_x^2\parder{u_{xx}} -\cdots$
denotes the (full) Euler operator restricted to solutions; 
$E_u^{(1,t)} = \parder{u_t} -\hat D_x\parder{u_{tx}} -2D_t\parder{u_{tt}} 
+ \hat D_x^2\parder{u_{txx}} + 2D_t\hat D_x\parder{u_{ttx}} 
+3D_t^2\parder{u_{ttt}} -\cdots$
and 
$E_u^{(1,x)} = \parder{u_x} -D_t\parder{u_{tx}} -2\hat D_x\parder{u_{xx}} 
+ D_t^2\parder{u_{ttx}} + 2D_t\hat D_x\parder{u_{txx}} 
+3\hat D_x^2\parder{u_{xxx}} -\cdots$
denote higher Euler operators restricted to solutions. 

Hence, the determining equations for multipliers 
$\wtil Q(t,x,u,u_t,u_x,u_{tx},u_{xx},u_{tt})$ 
are comprised by equations \eqref{adjsymmdeteq} and \eqref{Helmholtzalteq1}--\eqref{Helmholtzalteq4}. 
Computationally, 
these equations can be solved in a direct way 
after they are further split into a linear overdetermined system 
arising from the coefficients of $u_{txx}$, $u_{ttx}$, $u_{ttt}$, $u_{tttx}$.
This leads to the following classification result. 

\begin{thm}\label{class-multipliers}
(i) The gKN equation \eqref{gKN} admits no conservation law multipliers 
$\wtil Q(t,x,u)$ of differential order zero 
and no conservation law multipliers 
$\wtil Q(t,x,u,u_t,u_x)$ of differential order one, for any $f(u)$. 
\newline
(ii) For arbitrary $f(u)$, the gKN equation \eqref{gKN} admits
a single conservation law multiplier of mixed differential order two:
\begin{equation}
\wtil Q_1=\frac{u_{tx}u_x-u_{xx}u_t}{u_x^3} . 
\label{Q-1}
\end{equation}
\newline 
(iii) Additional conservation law multipliers of mixed differential order two 
are admitted by the gKN equation \eqref{gKN} 
only for the following $f(u)\not\equiv 0$:
\begin{align}
{\rm (a)}\qquad
& f(u)=C_1u^4+C_2u^3+C_3u^2+C_4u+C_5 \neq \pm(\tilde C_1 u^2 + \tilde C_2 u+\tilde C_3)^2
\nonumber\\
& \begin{aligned}
\wtil Q_{2\rm a}= & 
\frac{u_{tx}u_t-u_{tt}u_x}{u_x^3} 
-\frac{4}{3}\Big( \frac{u_{xx}^3}{u_x^6} + \frac{u_{xx}u_t}{u_x^5} + \frac{u_{tx}}{u_x^4}\Big) f(u)
+\frac{8}{9}\frac{u_{xx}}{u_x^6} f(u)^2
\\&\qquad
+2\Big( \frac{u_{xx}^2}{u_x^4} + \frac{2}{3} \frac{u_t}{u_x^3}\Big) f'(u)
-\frac{4}{3}\frac{u_{xx}}{u_x^2} f''(u) -\frac{4}{9}\frac{1}{u_x^4} f(u)f'(u) 
+ \frac{4}{9}f'''(u) 
\end{aligned}
\label{Q-2a}\\
& 
\wtil Q_{3\rm a}= 3t\wtil Q_{2\rm a} -x\wtil Q_1+\frac{4}{3}\frac{f(u)}{u_x^3} -2\frac{u_t}{u_x^2}
\label{Q-3a}\\
\nonumber\\
{\rm (b)}\qquad
& f(u)=\pm g(u)^2,
\quad 
g(u)=C_1 u^2+C_2 u+C_3
\nonumber\\
& 
\wtil Q_{2\rm b}=-2\frac{u_{xx}g(u)}{u_x^3}+2\frac{g'(u)}{u_x}
\label{Q-2b}\\
& 
\wtil Q_{3\rm b}=\wtil Q_{2\rm a}\big|_{f(u)=\pm g(u)^2},
\quad
\wtil Q_{4\rm b}=\wtil Q_{3\rm a}\big|_{f(u)=\pm g(u)^2}
\label{Q-3b4b}
\end{align}
\end{thm} 

\begin{rem} $f(u)$ is the square of a quadratic polynomial iff 
$4f(u)^2 f'''(u)-6f(u)f'(u)f''(u)+3f'(u)^3 =0$. 
\end{rem}

Each multiplier $\wtil Q$ determines 
a conserved density $\wtil T$ 
and a flux $\wtil X$, 
up to equivalence. 
The simplest way to obtain explicit expressions for them is by 
first splitting the characteristic equation \eqref{charalteqn}
with respect to $u_{xxx}$, $u_{txx}$, $u_{ttx}$, 
and next integrating the resulting linear system 
\begin{gather}
\wtil X_{u_{xx}}+ \wtil Q =0,
\quad
\wtil X_{u_{tt}}+ \wtil T_{u_{tx}}=0,
\quad
\wtil X_{u_{tx}} + \wtil T_{u_{xx}} =0,
\\
\begin{aligned}
& 
\wtil T_{t} + u_t \wtil T_{u} + u_{tt} \wtil T_{u_t} + u_{tx} \wtil T_{u_x}
+ \wtil X_{x} + u_x \wtil X_{u} + u_{tx} \wtil X_{u_t} + u_{xx} \wtil X_{u_x}
\\&\qquad
= \Big(u_t +\frac{3}{2}\frac{u_{xx}^2}{u_x} -\frac{f(u)}{u_x}\Big)\wtil Q . 
\end{aligned}
\end{gather}
This yields the following 
conserved densities $\wtil T(t,x,u,u_t,u_x,u_{tx},u_{xx})$
and fluxes $\wtil X(t,x,u,u_t,u_x,u_{tx},u_{xx},u_{tt})$:
\begin{align}
& 
\wtil T_1 = \frac{1}{2}\frac{u_{xx}^2}{u_x^2} +\frac{1}{3}\frac{f(u)}{u_x^2}
\label{T1}\\
& 
\wtil X_1 = \frac{1}{2}\frac{u_{xx}^2u_t}{u_x^3} -\frac{u_{xx}u_{tx}}{u_x^2} +\frac{1}{2}\frac{u_t^2}{u_x^2} -\frac{1}{3}\frac{u_t f(u)}{u_x^3}
\label{X1}\\
\nonumber\\
& 
\begin{aligned}
\wtil T_{2\rm a} = & 
\frac{1}{2}\frac{u_{xx}^2u_t}{u_x^3} -\frac{u_{xx}u_{tx}}{u_x^2} 
-\frac{1}{2}\frac{u_t^2}{u_x^2} - \frac{2}{3}\frac{u_{xx}^2f(u)}{u_x^4} 
+\frac{u_tf(u)}{u_x^3} -\frac{4}{9}\frac{f(u)^2}{u_x^4} +\frac{4}{9}f''(u)
\end{aligned}
\label{T2a}\\
&
\begin{aligned}
\wtil X_{2\rm a} = & 
\frac{1}{2}\frac{u_{tx}^2}{u_x^2} -\frac{u_{xx}u_{tx}u_t}{u_x^3} +\frac{u_{xx}u_{tt}}{u_x^2} 
+ \frac{1}{3}\frac{u_{xx}^4 f(u)}{u_x^6} + \frac{2}{3}\frac{u_{xx}^2u_t f(u)}{u_x^5}
\\&\qquad
+ \frac{4}{3}\frac{u_{xx}u_{tx} f(u)}{u_x^4} -\frac{4}{9}\frac{u_{xx}^2 f(u)^2}{u_x^6} 
-\frac{2}{3}\frac{u_{xx}^3 f'(u)}{u_x^4} -\frac{4}{3}\frac{u_{xx} u_t f'(u)}{u_x^3}
\\&\qquad 
+\frac{2}{3}\frac{u_{xx}^2 f''(u)}{u_x^2} +\frac{4}{9}\frac{u_{xx} f(u)f'(u)}{u_x^4} 
+\frac{1}{3}\frac{u_t^2 f(u)}{u_x^4} -\frac{4}{9}\frac{u_t f(u)^2}{u_x^5} 
\\&\qquad 
+\frac{4}{27}\frac{f(u)^3}{u_x^6} +\frac{2}{9}\frac{f'(u)^2-2f(u)f''(u)}{u_x^2}
-\frac{4}{9}u_{xx} f'''(u) +\frac{2}{9}u_x^2 f''''(u) 
\end{aligned}
\label{X2a}\\
\nonumber\\
& 
\wtil T_{3\rm a}=3t\wtil T_{2\rm a}-x\wtil T_1
\label{T3a}\\
& 
\wtil X_{3\rm a}=3t\wtil X_{2\rm a}-x\wtil X_1 -\frac{4}{3}\frac{u_{xx}f(u)}{u_x^3} +2\frac{u_{xx}u_t}{u_x^2} -\frac{4}{3}\frac{f'(u)}{u_x}
\label{X3a}\\
\nonumber\\
& 
\wtil T_{2\rm b}=\frac{g(u)}{u_x}
\label{T2b}\\
& 
\wtil X_{2\rm b}=\frac{u_{xx}^2 g(u)}{u_x^3} -2\frac{u_{xx}g'(u)}{u_x} +\frac{u_tg(u)}{u_x^2} \mp\frac{2}{3}\frac{g(u)^3}{u_x^3} +2u_x g''(u)
\label{X2b}\\
&
\begin{aligned}
\wtil T_{3\rm b} = & 
\frac{1}{2}\frac{u_{xx}^2u_t}{u_x^3} -\frac{u_{xx}u_{tx}}{u_x^2} 
-\frac{1}{2}\frac{u_t^2}{u_x^2} \mp\frac{2}{3}\frac{u_{xx}^2 g(u)^2}{u_x^4} 
\pm\frac{u_t g(u)^2}{u_x^3} -\frac{4}{9}\frac{g(u)^4}{u_x^4} 
\\&\qquad
\pm\frac{8}{9}\big(g(u)g''(u)+g'(u)^2\big)
\end{aligned}
\label{T3b}\\
&
\begin{aligned}
\wtil X_{3\rm b} = & 
\frac{1}{2}\frac{u_{tx}^2}{u_x^2} -\frac{u_{xx}u_{tx}u_t}{u_x^3} +\frac{u_{xx}u_{tt}}{u_x^2} 
\pm\frac{1}{3}\frac{u_{xx}^4 g(u)^2}{u_x^6} \pm\frac{2}{3}\frac{u_{xx}^2u_t g(u)^2}{u_x^5}
\\&\qquad
\pm\frac{4}{3}\frac{u_{xx}u_{tx} g(u)^2}{u_x^4} -\frac{4}{9}\frac{u_{xx}^2 g(u)^4}{u_x^6} 
\mp\frac{4}{3}\frac{u_{xx}^3 g(u)g'(u)}{u_x^4} \mp\frac{8}{3}\frac{u_{xx} u_t g(u)g'(u)}{u_x^3}
\\&\qquad 
\pm\frac{4}{3}\frac{u_{xx}^2\big(g(u) g''(u)+g'(u)^2\big)}{u_x^2} +\frac{8}{9}\frac{u_{xx} g(u)^3 g'(u)}{u_x^4} 
\pm\frac{1}{3}\frac{u_t^2 g(u)^2}{u_x^4} -\frac{4}{9}\frac{u_t g(u)^4}{u_x^5} 
\\&\qquad 
+\frac{4}{27}\frac{g(u)^6}{u_x^6} -\frac{8}{9}\frac{g(u)^3 g''(u)}{u_x^2}
\mp\frac{8}{9}u_{xx}\big(g(u)g'''(u)+3g'(u)g''(u)\big) 
\\&\qquad 
\pm\frac{4}{9}u_x^2\big(g(u) g''''(u) +4g'(u)g'''(u) +3g''(u)^2\big)
\end{aligned}
\label{X3b}\\
\nonumber\\
& 
\wtil T_{4\rm b}=3t\wtil T_{3\rm b}-x\wtil T_1
\label{T4b}\\
& 
\wtil X_{4\rm b}=3t\wtil X_{3\rm b}-x\wtil X_1 +2\frac{u_{xx}u_t}{u_x^2} 
\mp\frac{4}{3}\Big(\frac{u_{xx}g(u)^2}{u_x^3} +2\frac{g(u)g'(u)}{u_x}\Big)
\label{X4b}
\end{align}

We now rewrite these conserved densities and fluxes by 
first eliminating $u_t$, $u_{tx}$, $u_{tt}$ through the gKN equation \eqref{gKN}
and its differential consequences, 
and next adding appropriate total derivative terms 
$D_x \Theta(t,x,u,u_x,u_{xx},u_{xxx})$ and $-D_t \Theta(t,x,u,u_x,u_{xx},u_{xxx})$ 
to obtain equivalent conserved densities and fluxes that have
minimal differential order in $x$ derivatives. 
The results provide an explicit classification of 
all conservation laws \eqref{utelimconslaw} up to differential order three
(in $x$ derivatives). 

\begin{thm}\label{class-conslaws}
The gKN equation \eqref{gKN} admits:
\newline
(i) 
no conservation laws \eqref{chareqn} of differential order zero, 
for any $f(u)$. 
\newline
(ii) 
a single conservation law \eqref{chareqn} of differential order one
\begin{align}
&
T_{2\rm b}=\frac{g(u)}{u_x}
\label{1stordT-2b}\\
& 
X_{2\rm b}=\frac{u_{xxx} g(u)}{u_x^2} -\frac{1}{2}\frac{u_{xx}^2 g(u)}{u_x^3} 
-2\frac{u_{xx}g'(u)}{u_x} \pm\frac{1}{3}\frac{g(u)^3}{u_x^3} +2u_x g''(u)
\label{1stordX-2b}
\end{align}
with 
\begin{equation}
f(u)=\pm g(u)^2,
\quad
g'''(u)=0
\quad 
(g(u)=C_1 u^2 + C_2 u+C_3) . 
\label{1stord-f}
\end{equation}
(iii) 
a single conservation law \eqref{chareqn} of differential order two:
\begin{align}
& 
T_1 = \frac{1}{2}\frac{u_{xx}^2}{u_x^2} +\frac{1}{3}\frac{f(u)}{u_x^2}
\label{2ndordT-1}\\
& 
\begin{aligned}
X_1 = & -\frac{u_{xxxx}u_{xx}}{u_x^2} +\frac{1}{2}\frac{u_{xxx}^2}{u_x^2} 
+\frac{2}{3}\frac{u_{xxx}(3u_{xx}^2+f(u))}{u_x^3} 
-\frac{9}{8}\frac{u_{xx}^4}{u_x^4} 
\\&\qquad
+\frac{1}{2}\frac{u_{xx}^2 f(u)}{u_x^4} 
-\frac{u_{xx}f'(u)}{u_x^2} +\frac{1}{6}\frac{f(u)^2}{u_x^4} 
\end{aligned}
\label{2ndordX-1}
\end{align}
with
\begin{equation}
f(u) \text{ arbitrary } .
\end{equation}
(iv) 
two conservation laws \eqref{chareqn} of differential order three:
\begin{align}
& 
T_{2\rm a} = 
\frac{1}{2}\frac{u_{xxx}^2}{u_x^2} -\frac{3}{8}\frac{u_{xx}^4}{u_x^4} 
+\frac{5}{6}\frac{u_{xx}^2 f(u)}{u_x^4} +\frac{1}{18}\frac{f(u)^2}{u_x^4} 
-\frac{5}{9}f''(u)
\label{3rdordT-2a}\\
&
\begin{aligned}
X_{2\rm a} = & 
-\frac{u_{xxxxx}u_{xxx}}{u_x^2} +\frac{1}{2}\frac{u_{xxxx}^2}{u_x^2} 
+\frac{u_{xxxx}u_{xxx}u_{xx}}{u_x^3} 
+\frac{1}{6}\frac{u_{xxxx}u_{xx}(9u_{xx}^2-10f(u))}{u_x^4} 
\\&\qquad
+ 3\frac{u_{xxx}^3}{u_x^3} 
-\frac{2}{3} \frac{u_{xxx}^2(9u_{xx}^2 -2f(u))}{u_x^4} 
+ \frac{1}{9}\frac{u_{xxx}(15u_{xx}^2 +2f(u))f(u)}{u_x^5} 
\\&\qquad
+\frac{5}{3}\frac{u_{xxx}u_{xx} f'(u)}{u_x^3} 
+\frac{9}{8}\frac{u_{xx}^6}{u_x^6} 
- \frac{29}{12}\frac{u_{xx}^4 f(u)}{u_x^6} 
-\frac{1}{6}\frac{u_{xx}^3 f'(u)}{u_x^4} 
\\&\qquad 
+\frac{19}{18}\frac{u_{xx}^2 f(u)^2}{u_x^6} 
-\frac{5}{6}\frac{u_{xx}^2 f''(u)}{u_x^2} 
-\frac{5}{9}\frac{u_{xx} f(u)f'(u)}{u_x^4} 
-\frac{4}{9} u_{xx} f'''(u) 
\\&\qquad 
+\frac{1}{27}\frac{f(u)^3}{u_x^6} 
+\frac{5}{18}\frac{2f(u)f''(u)-f'(u)^2}{u_x^2}
+\frac{2}{9}u_x^2 f''''(u) 
\end{aligned}
\label{3rdordX-2a}
\end{align}
and 
\begin{align}
& 
T_{3\rm a}=3t T_{2\rm a}-x T_1
\label{3rdordT-3a}\\
& 
X_{3\rm a}=3t X_{2\rm a}-x X_1 -\frac{u_{xxx}u_{xx}}{u_x^2} 
+\frac{2}{3}\frac{u_{xx}f(u)}{u_x^3} +\frac{5}{3}\frac{f'(u)}{u_x}
\label{3rdordX-3a}
\end{align}
both with 
\begin{equation}
f'''''(u) =0 
\quad 
(f(u)=C_1u^4+C_2u^3+C_3u^2+C_4u+C_5) .
\label{3rdord-f}
\end{equation}
\end{thm} 

The multipliers \eqref{TQrelation} corresponding to 
these conservation laws \eqref{1stordT-2b}--\eqref{3rdordX-3a}
are found to coincide with the multipliers \eqref{Q-1}--\eqref{Q-3b4b} 
of mixed differential order two 
when $u_{xxx}$, $u_{xxxx}$, $u_{xxxxx}$ are eliminated through the gKN equation 
in the solved form \eqref{u3x}. 
From this correspondence, 
we remark that the conserved densities and fluxes in Corollary~\ref{class-conslaws}
can be also obtained directly from the multipliers 
$\wtil Q(t,x,u,u_t,u_x,u_{tx},u_{xx},u_{tt})$ 
classified in Theorem~\ref{class-multipliers}
either by using a homotopy integral formula
or by applying a scaling formula based on the scaling transformation \eqref{scaling}, 
as follows. 

Consider the homotopy 
\begin{equation}
u_{(\lambda)}=\lambda u + (1-\lambda)x
\end{equation}
where we have chosen $u_{(0)}=x$ so that, for each multiplier, 
$\wtil Q(t,x,u_{(0)},u_{(0)t},u_{(0)x},u_{(0)tx},$ $u_{(0)xx},u_{(0)tt})
= \wtil Q(t,x,x,0,1,0,0,0)$ is a non-singular function of $t,x$. 
Then, using the general method from Refs.\cite{AncBlu02a,AncBlu02b},
we can show that the integral formula 
\begin{equation}\label{TfromQ-homotopy}
T=(u-x)\int_0^1 \wtil Q(t,x,u_{(\lambda)},u_{(\lambda)t},u_{(\lambda)x},u_{(\lambda)tx},u_{(\lambda)xx},u_{(\lambda)tt}) d\lambda\big|_{u_t =u_{xxx} - (\frac{3}{2}u_{xx}^2 +f(u))/u_x} 
\end{equation}
holds up to equivalence. 
When applied to the multipliers \eqref{Q-1}--\eqref{Q-3b4b} from Theorem~\ref{class-multipliers}, 
this formula is readily checked to reproduce the conserved densities \eqref{1stordT-2b}, \eqref{2ndordT-1}, \eqref{3rdordT-2a}, \eqref{3rdordT-3a}. 

Alternatively, consider the scaling equivalence transformation \eqref{scaling}, 
under which the form of the gKN equation \eqref{gKN} is preserved. 
Each multiplier is manifestly homogeneous under this transformation, 
\begin{equation}
\wtil Q\rightarrow\lambda^\omega \wtil Q
\end{equation}
which implies 
\begin{equation}
T\rightarrow\lambda^\omega T,
\quad
X\rightarrow\lambda^{\omega-2} X
\end{equation}
where $\omega$ denotes the scaling weight of the multiplier. 
Since $u$ appears in each multiplier only through $f(u)$, $f'(u)$, and so on, 
it is now convenient to write $v=u_x$, $v_x=u_{xx}$, and so on, 
and replace $f'(u)=v^{-1}D_xf$, and so on. 
Thus, we have 
\begin{equation}
\begin{aligned}
& \wtil Q(t,x,u,u_t,u_x,u_{tx},u_{xx},u_{tt})\big|_{u_t =u_{xxx} - (\frac{3}{2}u_{xx}^2 +f(u))/u_x}
\\&\qquad
= \hat{Q}(t,x,v,v_x,v_{xx},v_{xxx},v_{xxxx},v_{xxxxx},f,D_xf,D_x^2f)
\end{aligned}
\end{equation}
and similarly
\begin{equation}
T(t,x,u,u_x,u_{xx},u_{xxx}) 
= \hat{T}(t,x,v,v_x,v_{xx},f,D_xf,D_x^2f) .
\end{equation}
By 
using the infinitesimal generator $\X_{\text{scal.}}=x\parder{x}+3t\parder{t}-4f\parder{f}$ 
of the scaling transformation \eqref{scaling}, 
we derive the relation 
\begin{equation}
\begin{aligned}
\pr\X_{\text{scal.}}\hat{T} 
& = \omega\hat{T} 
\\
& = xD_x\hat{T} + 3tD_t\hat{T}
-(v+xv_x+3tv_t) \hat{T}_{v} - (4f+xD_xf)\hat{T}_f 
\\&\qquad
- (2v_x+xv_{xx}+3tv_{tx}) \hat{T}_{v_x} 
- (5D_xf+xD_x^2f)\hat{T}_{D_xf}
\\&\qquad
- (3v_{xx}+xv_{xxx}+3tv_{txx}) \hat{T}_{v_{xx}} 
- (6D_x^2f+xD_x^3f) \hat{T}_{D_x^2f} . 
\end{aligned}
\end{equation}
We next expand out the term $3tD_t\hat{T}$ 
and integrate by parts with respect to $x$ in all of the remaining terms. 
This yields
\begin{equation}
(\omega+1)\hat{T} = 
3t\hat{T}_t - (4f+xD_xf) E_f(\hat{T}) - (v+xv_x) E_v(\hat{T}) + D_x\Theta
\end{equation}
where $E_f,E_v$ denote the (spatial) Euler operators with respect to the variables $f,v$.
After expressing $v+xv_x=D_x(xv)$ and $f+xD_xf=D_x(xf)$, 
we integrate by parts again 
and simplify the remaining terms that contain $x$ by using the identity
$x w = xD_xD_x^{-1}w = D_x(xD_x^{-1}w) - D_x^{-1}w$. 
Hence we get 
\begin{equation}\label{EulerTeqn1}
(\omega+1)\hat{T} = 
3t\hat{T}_t - 3f E_f(\hat{T}) -D_x^{-1}( fD_xE_f(\hat{T}) +vD_xE_v(\hat{T}) ) + D_x\wtil\Theta
\end{equation}
Finally, we observe 
\begin{equation}\label{EulerTeqn2}
\hat{Q}=E_u(\hat{T})= f'Q^f -D_xQ^v, 
\quad
Q^f =E_f(\hat{T}),
\quad
Q^v= E_v(\hat{T})
\end{equation}
due to the variational identity 
\begin{equation}\label{varid}
\frac{\delta}{\delta u} = -D_x\frac{\delta}{\delta v} + f'(u)\frac{\delta}{\delta f} .
\end{equation}
Combining equations \eqref{EulerTeqn1} and \eqref{EulerTeqn2}, 
we now have an explicit formula 
\begin{equation}\label{TfromQ-scaling}
(\omega+1)\hat{T} 
= 3t\hat{T}_t - 3f \hat{Q}^f -D_x^{-1}( fD_x\hat{Q}^f+vD_x\hat{Q}^v ) 
= 3t\hat{T}_t - 4f \hat{Q}^f +D_x^{-1}( vD_x\hat{Q} ) 
\end{equation}
which yields, up to equivalence, $\hat{T}$ in terms of $\hat{Q}$
if $\omega+1\neq 0$. 
For the multipliers \eqref{Q-1}--\eqref{Q-3b4b}, 
we find their scaling weights are given by 
$\omega_1=-2$, $\omega_{2\rm a}=\omega_{3\rm b}=-4$, $\omega_{3\rm a}=\omega_{4\rm b}=-1$, $\omega_{2\rm b}=-1$.
Thus, 
as $\omega_1+1=-1$ and $\omega_{2\rm a}+1=\omega_{3\rm b}+1=-3$ are non-zero,
the formula \eqref{TfromQ-scaling} can be applied to the multipliers
\eqref{Q-1} and \eqref{Q-2a}. 
It is straightforward to check that this reproduces
the conserved densities \eqref{2ndordT-1} and \eqref{3rdordT-2a}. 

The conserved density \eqref{2ndordT-1} agrees with the lowest-order Hamiltonian
for the KN equation \cite{Sok84,Mok91}. 
Since this conserved density is also admitted by 
the gKN equation for arbitrary $f(u)$, 
we conclude that the basic Hamiltonian structure of the KN equation 
carries over to the gKN equation
\begin{equation}\label{gKNHamilstruc}
u_t = u_{xxx} - \frac{3}{2}\frac{u_{xx}^2}{u_x} + \frac{f(u)}{u_x}
= \Hop\big( \delta T_1/\delta u \big)
\end{equation}
where
\begin{equation}\label{Hamilop}
\Hop= u_x D_x^{-1}u_x
\end{equation}
is a Hamiltonian operator. 
In the case of the KN equation itself, 
there is a hierarchy of conserved densities generated by the adjoint of 
the fourth order recursion operator of the KN equation \cite{DemSok08}. 
The hierarchy is produced by acting with this adjoint operator on multipliers,
starting from the multiplier $Q=E_u(T_1)$ corresponding to 
the conserved density $T_1$. 
A second hierarchy of conserved densities arises similarly from 
the multiplier corresponding to the conserved density \eqref{3rdordT-2a}. 
All of these densities are invariant under translations on $t,x$. 

The conserved density \eqref{3rdordT-3a} that explicitly involves $t,x$
is new and does not seem to be connected with a recursion operator for 
the KN equation. 
The other conserved density \eqref{1stordT-2b} is also new, 
but it is admitted only in the quadratic case of the KN equation. 

The new conserved density \eqref{3rdordT-3a} 
can be expressed in a simple form in terms of the first two 
local Hamiltonians of the KN equation \eqref{KNeq}:
\begin{equation}\label{newT}
T= x H_{(1)} -3t H_{(2)}
\end{equation}
where
\begin{gather}
H_{(1)}=\frac{1}{2}\frac{u_{xx}^2}{u_x^2} +\frac{1}{3}\frac{p(u)}{u_x^2} 
=T_1|_{f(u)=p(u)}
\label{H1}\\
H_{(2)}= \frac{1}{2}\frac{u_{xxx}^2}{u_x^2} -\frac{3}{8}\frac{u_{xx}^4}{u_x^4} 
+\frac{5}{6}\frac{u_{xx}^2 p(u)}{u_x^4} +\frac{1}{18}\frac{p(u)^2}{u_x^4} +\frac{5}{9}p''(u) 
=T_{2\rm a}|_{f(u)=p(u)}
\label{H2}
\end{gather}
with $p(u)$ being an arbitrary quartic polynomial \eqref{quarticKN}. 
We can interpret expression \eqref{newT} 
like the conserved density involving $t,x$ for the KdV equation,
which is related to Galilean invariance and motion of center of mass
for KdV solutions. 

Consider the conserved quantities \eqref{C} defined from 
the Hamiltonian densities \eqref{H1} and \eqref{H2} evaluated 
for solutions of the KN equation \eqref{KNeq}. 
The first Hamiltonian density yields the Hamiltonian energy 
\begin{equation}
\mathcal E_1 =\int^\infty_{-\infty} H_{(1)} dx
\end{equation}
which is a constant of the motion for all solutions $u(t,x)$ of the KN equation
having sufficiently rapid spatial decay as $x\rightarrow \pm\infty$. 
Similarly the second Hamiltonian density yields a higher-order energy 
\begin{equation}
\mathcal E_2 =\int^\infty_{-\infty} H_{(2)} dx . 
\end{equation}
The conserved quantity defined from the new density \eqref{newT} is then 
given by 
\begin{equation}
\mathcal C = \int^\infty_{-\infty} (xH_{(1)}-3tH_{(2)})dx
= \mathcal X_1(t)-3t\mathcal E_2 
\end{equation}
where 
\begin{equation}
\mathcal X_1(t) = \int^\infty_{-\infty} xH_{(1)} dx
\end{equation}
is the center of energy (or first $x$-moment of energy) for solutions $u(t,x)$.
Since $\mathcal C$ is a constant of motion 
for solutions $u(t,x)$ of the KN equation 
that have sufficiently rapid spatial decay as $x\rightarrow \pm\infty$, 
we conclude 
$\mathcal C = \mathcal C|_{t=0} = \mathcal X_1(0)$.
This yields the relation 
\begin{equation}
\mathcal X_1(t) = 3t\mathcal E_2 + \mathcal X_1(0)
\end{equation}
which expresses the property that the center of energy 
of solutions moves at a constant speed (equal to $3\mathcal E_2$).

\section{Action of point symmetries on conserved densities}
\label{symmaction}

Point symmetries have a natural action on conservation laws. 
In terms of the infinitesimal generator \eqref{pointsymm} of a point symmetry,
its action on the conserved density $T$ and flux $X$ 
in a conservation law \eqref{conslaw} of the gKN equation \eqref{gKN}
is given by \cite{BluTemAnc}
\begin{equation}\label{symmTX}
\begin{aligned}
& \wtil T = \pr\X(T) +(D_x\xi+D_t\tau)T -TD_t\tau -XD_x\tau 
\\
& \wtil X = \pr\X(X) +(D_x\xi+D_t\tau)X -TD_t\xi -XD_x\xi . 
\end{aligned}
\end{equation}
This action produces a conserved density $\wtil T$ and a flux $\wtil X$, 
satisfying $D_t\wtil T + D_x\wtil X=0$
for all solutions $u(t,x)$ of the gKN equation \eqref{gKN}. 
Through the variational relation \eqref{TQrelation}, 
it is straightforward to show that the corresponding point symmetry action 
on multipliers is given by 
\begin{equation}\label{symmQ}
\wtil Q=E_u(\wtil T) = E_u((\eta-\tau u_t-\xi u_x)Q),
\quad
Q=E_u(T) . 
\end{equation}
This formula \eqref{symmQ} can be used to determine 
the transformed conserved density $\wtil T$, 
since there is a one-to-one correspondence between 
multipliers and conserved densities modulo trivial densities. 
In particular, 
the transformed conserved density $\wtil T$ is trivial \eqref{trivconslaw} 
if (and only if) $\wtil Q =0$ holds identically. 

From Theorem~\ref{class-pointsymm}, 
the point symmetries admitted by the gKN equation 
fall into two different classes. 
The first class holds for arbitrary $f(u)$ and consists only of 
translations \eqref{pointsymm-general} on $t,x$. 
The second class comprises six different forms 
\eqref{pointsymm-a}--\eqref{pointsymm-f} of $f(u)$, 
as derived from the ODE \eqref{fODE}. 
Similarly, from Corollary~\ref{class-conslaws}, 
the conserved densities admitted by the gKN equation 
fall into three classes,
one holding when $f(u)$ is arbitrary, 
another holding when $f(u)$ is a general quartic polynomial \eqref{3rdord-f}, 
and the other holding when $f(u)$ is the square of a general quadratic polynomial \eqref{1stord-f}. 

To begin, we specialize the point symmetries 
in the class \eqref{pointsymm-a}--\eqref{pointsymm-f} 
to the cases when $f(u)$ is either a general quartic polynomial
or the square of a quadratic polynomial. 

For the first case, we have:
\begin{align}
{\rm (a')}\qquad
\label{pointsymm-a-quartic}
&\begin{aligned}
& f(u)=C (u+a)^{k},
\quad
k=0,1,2,3
\\
& \X_{3\rm a'}=\tfrac{3}{4}(2-k)t\parder{t}+\tfrac{1}{4}(2-k)x\parder{x} + (u+a)\parder{u}
\end{aligned}
\\\nonumber\\
{\rm (b')}\qquad
\label{pointsymm-b-quartic}
&\begin{aligned}
& f(u)=C (u+a)^4
\\
& \X_{3\rm b'}=-(3t/2)\parder{t}-(x/2)\parder{x} + (u+a)\parder{u} ,
\\
& \X_{4\rm b'}=(u+a)^2\parder{u} 
\end{aligned}
\\\nonumber\\
{\rm (c')}\qquad
\label{pointsymm-c-quartic}
&\begin{aligned}
& f(u)=C(a+b+u)^{k}(a-b-u)^{4-k},
\quad
a\neq0 ,
\quad
k=1,2
\\
& \X_{3\rm c'}=(3a(k-2)t/2)\parder{t}+(a(k-2)x/2)\parder{x} + ((u+b)^2-a^2)\parder{u}
\end{aligned}
\end{align}
Note class $(\rm c')$ is preserved under $k\rightarrow 4-k$, $a\leftrightarrow -a$. 

For the second case, we write $f(u)=g(u)^2$. 
Then we have:
\begin{align}
{\rm (a'')}\qquad
\label{pointsymm-a-quadrsqr}
&\begin{aligned}
& g(u)=C (u+a)
\\
& \X_{3\rm a''}=X_{3\rm a'}|_{k=2}= (u+a)\parder{u}
\end{aligned}
\\\nonumber\\
{\rm (b'')}\qquad
\label{pointsymm-b-quadrsqr}
&\begin{aligned}
& g(u)=C (u+a)^2
\\
& \X_{3\rm b''} = X_{3\rm b'} =-(3t/2)\parder{t}-(x/2)\parder{x} + (u+a)\parder{u} 
\\
& \X_{4\rm b''} = X_{4\rm b'} = (u+a)^2\parder{u} 
\end{aligned}
\\\nonumber\\
{\rm (c'')}\qquad
\label{pointsymm-c-quadrsqr}
&\begin{aligned}
& g(u)=C(a^2-(b+u)^2) , 
\quad
a\neq0 
\\
& \X_{3\rm c''}=X_{3\rm c'}|_{k=2} = ((u+b)^2-a^2)\parder{u}
\end{aligned}
\end{align}

By applying the formula \eqref{symmQ}, 
we now obtain the results summarized in Tables~\ref{T1symmaction}--\ref{T2bsymmaction}. 

\begin{table}[ht!]
\begin{center}
\begin{tabular}{|c|c|c|c|c|c|c|c|c|c|}
\hline
\hfill  & $\X_1$ & $\X_2$ & $\X_{\rm 3a}$ & $\X_{\rm 3b}$ & $\X_{\rm 3c}$ & $\X_{\rm 3d}$ & $\X_{\rm 3e}$ & $\X_{\rm 3f}$ & $\X_{\rm 4f}$ \\
\hline
$T_1$
& $0$
& $0$ 
& $aT_1$ 
& $-aT_1$ 
& $-aT_1$ 
& $-aT_1$ 
& $aT_1$ 
& $\tfrac{1}{2}T_1$ 
& $0$
\\
\hline\hline
$f(u)$
& arb.
& arb.
& \eqref{pointsymm-a}
& \eqref{pointsymm-b}
& \eqref{pointsymm-c}
& \eqref{pointsymm-d}
& \eqref{pointsymm-e}
& \eqref{pointsymm-f}
& \eqref{pointsymm-f}
\\
\hline
\end{tabular}
\end{center}
\caption{Action of point symmetries on conserved density~\eqref{2ndordT-1} of the gKN equation~\eqref{gKN}}
\label{T1symmaction}
\end{table}

\begin{table}[ht!]
\begin{center}
\begin{tabular}{|c|c|c|c|c|c|c|}
\hline
\hfill  & $\X_1$ & $\X_2$ & $\X_{\rm 3a'}$ & $\X_{\rm 3b'}$ & $\X_{\rm 4b'}$ & $\X_{\rm 3c'}$ \\
\hline
$T_{\rm 2a}$
& $0$
& $0$ 
& $\tfrac{3}{4}(k-2)T_{\rm 2a}$
& $\tfrac{3}{2}T_{\rm 2a}$
& $0$
& $\tfrac{1}{2}a(2-k)T_{\rm 2a}$
\\
\hline
$T_{\rm 3a}$
& $3T_{\rm 2a}$
& $-T_1$
& $0$
& $0$
& $0$
& $0$
\\
\hline\hline
$f(u)$
& arb.
& arb.
& \eqref{pointsymm-a-quartic}
& \eqref{pointsymm-b-quartic}
& \eqref{pointsymm-b-quartic}
& \eqref{pointsymm-c-quartic}
\\
\hline
\end{tabular}
\end{center}
\caption{Action of point symmetries on conserved densities~\eqref{3rdordT-2a} 
and~\eqref{3rdordT-3a} of the gKN equation~\eqref{gKN}}
\label{T2aT3asymmaction}
\end{table}

\begin{table}[ht!]
\begin{center}
\begin{tabular}{|c|c|c|c|c|c|c|c|}
\hline
\hfill  & $\X_1$ & $\X_2$ & $\X_{\rm 3a''}$ & $\X_{\rm 3b''}$ & $\X_{\rm 4b''}$ & $\X_{\rm 3c''}$ \\
\hline
$T_{\rm 2b}$
& $0$
& $0$ 
& $0$
& $-T_{\rm 2b}$
& $0$
& $0$
\\
\hline\hline
$f(u)$
& arb.
& arb.
& \eqref{pointsymm-a-quadrsqr}
& \eqref{pointsymm-b-quadrsqr}
& \eqref{pointsymm-b-quadrsqr}
& \eqref{pointsymm-c-quadrsqr}
\\
\hline
\end{tabular}
\end{center}
\caption{Action of point symmetries on conserved density~\eqref{1stordT-2b} of the gKN equation~\eqref{gKN}}
\label{T2bsymmaction}
\end{table}

\section{Nonlocal symmetries and nonlocal conservation laws}
\label{nonlocal}

The Hamiltonian structure \eqref{gKNHamilstruc}--\eqref{Hamilop} of 
the gKN equation \eqref{gKN} gives rise to a natural mapping 
from conserved densities into symmetries of a special form as follows. 

\begin{prop}\label{Hamilmapping}
For the gKN equation \eqref{gKN}, 
$P=\Hop Q$ is the characteristic function of a symmetry
whenever $Q$ is an adjoint-symmetry (or multiplier),
where $\Hop$ is the Hamiltonian operator \eqref{Hamilop}. 
In particular, if $T$ is a conserved density, then 
$P=\Hop(\delta T/\delta u)$ is a Hamiltonian symmetry. 
\end{prop}

This mapping is a general result (see Ref.~\cite{Olv}) for Hamiltonian PDEs. 
Note that not every symmetry admitted by the gKN equation will have the form $P=\Hop Q$. 
Moreover, 
also note that $P=\Hop Q$ can produce a nonlocal symmetry 
due to the appearance of $D_x^{-1}$ in the Hamiltonian operator \eqref{Hamilop}
for the gKN equation. 

A function of $t$, $x$, $u$, and $x$-derivatives of $u$ is {\em nonlocal}
if it contains $D_x^{-1}$ applied to an expression or a variable that is not 
a total $x$-derivative of a local function of $t$, $x$, $u$, and finitely many $x$-derivatives of $u$. 

A {\em nonlocal symmetry} of the gKN equation \eqref{gKN}
is an infinitesimal generator $\hat\X=P\parder{u}$
where $P$ is a nonlocal function satisfying the symmetry determining equation \eqref{symmdeteq}. 
Similarly, 
a {\em nonlocal adjoint-symmetry} (or cosymmetry) of the gKN equation \eqref{gKN}
is a nonlocal function $Q$ that satisfies the adjoint-symmetry determining equation \eqref{adjsymmdeteq}. 

A conservation law $D_t T +D_x X=0$ holding for all solutions~$u(t,x)$ of the gKN equation
is {\em nonlocal} if the multiplier $Q=\delta T/\delta u$ is a nonlocal function, 
where $\delta/\delta u$ denotes the variational derivative defined by 
\begin{equation}
\int_{-\infty}^{\infty}\delta T\; dx = \int_{-\infty}^{\infty} (\delta T/\delta u)\delta u\; dx \quad\text{(modulo boundary terms)}
\end{equation}
for any variation $\delta u(t,x)$. 
When the conserved density is a local function $T(t,x,u,u_x,\ldots)$, 
note that $\delta T/\delta u=E_u(T)$ then coincides with
the relation \eqref{TQrelation}.

The classification of multipliers and conservation laws 
in Theorems~\ref{class-multipliers} and~\ref{class-conslaws}
for the gKN equation 
now leads directly to a classification of Hamiltonian symmetries 
by Proposition~\ref{Hamilmapping}, 
under the restriction that the Hamiltonian conserved density is local. 

\begin{thm}\label{class-Hamilsymms}
The gKN equation \eqref{gKN} admits:
\newline
(i) 
a single Hamiltonian symmetry 
having a local Hamiltonian of differential order at most one
\begin{equation}\label{P-2b}
\hat\X_{1}=g(u)\parder{u},
\quad
P_{1}=g(u)=\Hop\big(\delta T/\delta u\big),
\quad
T=\frac{g(u)}{u_x}
\end{equation}
with 
\begin{equation}
f(u)=\pm g(u)^2,
\quad
g'''(u)=0
\quad 
(g(u)=C_1 u^2 + C_2 u+C_3) . 
\end{equation}
(ii) 
a single Hamiltonian symmetry having a local Hamiltonian of differential order two
\begin{equation}\label{P-1}
\begin{aligned}
& 
\hat\X_{2}= \Big( u_{xxx} - \frac{3}{2}\frac{u_{xx}^2}{u_x} + \frac{f(u)}{u_x}\Big)\parder{u} = u_t \parder{u},
\\
&
P_{2}=u_t = \Hop\big(\delta T/\delta u\big),
\quad
T = H_{(1)}=\frac{1}{2}\frac{u_{xx}^2}{u_x^2} +\frac{1}{3}\frac{f(u)}{u_x^2}
\end{aligned}
\end{equation}
with
\begin{equation}
f(u) \text{ arbitrary.}
\end{equation}
(iii) 
two Hamiltonian symmetries having a local Hamiltonian of differential order three
\begin{equation}\label{P-2a}
\begin{aligned}
& 
\hat\X_{3\rm a} =-\Big( u_{xxxxx} -5\frac{u_{xxxx}u_{xx}}{u_x}
-\frac{5}{2}\frac{u_{xxx}^2}{u_x} +\frac{25}{2}\frac{u_{xxx}u_{xx}^2}{u_x^2} 
-\frac{45}{8}\frac{u_{xx}^4}{u_x^3} 
-\frac{5}{3}\frac{u_{xxx} f(u)}{u_x^2} 
\\&\qquad\qquad
+\frac{25}{6}\frac{u_{xx}^2 f(u)}{u_x^3} -\frac{5}{3}\frac{u_{xx} f'(u)}{u_x} 
-\frac{5}{18}\frac{f(u)^2}{u_x^3} +\frac{5}{9}u_x f''(u) \Big)\parder{u}
= -\Rop(u_x)\parder{u},
\\
&
P_{3\rm a} = -\Rop(u_x) = \Hop\big(\delta T/\delta u\big),
\\
&
T = H_{(2)}= 
\frac{1}{2}\frac{u_{xxx}^2}{u_x^2} -\frac{3}{8}\frac{u_{xx}^4}{u_x^4} 
+\frac{5}{6}\frac{u_{xx}^2f(u)}{u_x^4} 
+\frac{1}{18}\frac{f(u)^2}{u_x^4} -\frac{5}{9}f''(u)
\end{aligned}
\end{equation}
and 
\begin{equation}\label{P-3a}
\begin{aligned}
& 
\hat\X_{3\rm b} = \bigg( 
3t\Big( u_{xxxxx} -5\frac{u_{xxxx}u_{xx}}{u_x}
-\frac{5}{2}\frac{u_{xxx}^2}{u_x} +\frac{25}{2}\frac{u_{xxx}u_{xx}^2}{u_x^2} 
-\frac{45}{8}\frac{u_{xx}^4}{u_x^3} 
-\frac{5}{3}\frac{u_{xxx} f(u)}{u_x^2} 
\\&\qquad\qquad
+\frac{25}{6}\frac{u_{xx}^2 f(u)}{u_x^3} -\frac{5}{3}\frac{u_{xx} f'(u)}{u_x} 
-\frac{5}{18}\frac{f(u)^2}{u_x^3} +\frac{5}{9}u_x f''(u) \Big)
+x \Big( u_{xxx} - \frac{3}{2}\frac{u_{xx}^2}{u_x} + \frac{f(u)}{u_x} \Big)
\\&\qquad\qquad
+ u_{xx} - u_xD_x^{-1}\Big( \frac{1}{2}\frac{u_{xx}^2}{u_x^2} +\frac{1}{3}\frac{f(u)}{u_x^2} \Big) 
\bigg)\parder{u},
\\
& 
P_{3\rm b} = 3t \Rop(u_x) +x u_t + u_{xx} - u_xD_x^{-1} H_{(1)}
= \Hop\big(\delta T/\delta u\big),
\quad
T=-3t H_{(2)}+x H_{(1)}
\end{aligned}
\end{equation}
both with 
\begin{equation}
f'''''(u) =0 
\quad 
(f(u)=C_1u^4+C_2u^3+C_3u^2+C_4u+C_5) .
\end{equation}
\end{thm} 

Here 
\begin{equation}\label{Rop1}
\Rop = D_x^4 + a_1D_x^3+a_2D_x^2+a_3D_x +a_4 -P_{(1)}D_x^{-1}Q_{(1)} -P_{(0)}D_x^{-1}Q_{(2)}  
\end{equation}
is the known 4th order recursion operator of the KN equation \eqref{KNeq},
where the coefficients $a_1,\ldots,a_4$ 
are specific functions of $u$ and $x$-derivatives of $u$, 
shown in Ref.~\cite{DemSok08},
and where 
\begin{equation}\label{rootP0P1}
P_{(0)} = -u_x,
\quad
P_{(1)} = u_t = u_{xxx} - \frac{3}{2}\frac{u_{xx}^2}{u_x} + \frac{f(u)}{u_x}
\end{equation}
are the characteristic functions of $x$-translation and $t$-translation symmetries,
and 
\begin{gather}
Q_{(1)} = E_u(H_{(1)}) = 
\frac{u_{xxxx}}{u_x^2}
-4\frac{u_{xxx}u_{xx}}{u_x^3}
+3\frac{u_{xx}^3}{u_x^4}
-2\frac{u_{xx}f(u)}{u_x^4} 
+\frac{f'(u)}{u_x^2} 
\label{rootQ0}\\
\begin{aligned}
Q_{(2)} = E_u(H_{(2)}) = &
-\frac{u_{xxxxxx}}{u_x^2}
+6\frac{u_{xxxxx}u_{xx}}{u_x^3}
+10\frac{u_{xxxx}u_{xxx}}{u_x^3} 
-\frac{5}{6}\frac{u_{xxxx}(27u_{xx}^2-2f(u))}{u_x^4} 
\\&\qquad
=30\frac{u_{xxx}^2u_{xx}}{u_x^4} 
+\frac{20}{3}\frac{u_{xxx}u_{xx}(9u_{xx}^2-2f(u))}{u_x^5} 
+\frac{10}{3}\frac{u_{xxx}f'(u)}{u_x^3} 
\\&\qquad
-\frac{5}{6}\frac{u_{xx}^3(27u_{xx}^2-20f(u))}{u_x^6} 
-\frac{15}{2}\frac{u_{xx}^2f'(u))}{u_x^4} 
+\frac{5}{3}\frac{u_{xx}f''(u)}{u_x^2} 
\\&\qquad
+\frac{5}{9}\frac{f(u)f'(u)}{u_x^4} 
-\frac{5}{9}f'''(u)
\end{aligned}
\label{rootQ1}
\end{gather}
are the respective multipliers for the conserved densities $H_{(1)}$ and $H_{(2)}$,
with $f(u)$ denoting an arbitrary quartic polynomial \eqref{quarticKN}. 

This recursion operator can be used to obtain 
a hierarchy of symmetries for the KN equation.
Furthermore, 
a hierarchy of conserved densities can be obtained
if, as suggested in Ref.~\cite{DemSok08}, 
the operator $\Eop=\Rop\Hop^{-1}$ defines a Hamiltonian operator 
compatible with $\Hop$,
so that $\Eop$ and $\Hop$ are a bi-Hamiltonian pair \cite{Olv}. 
(A proof of this property for $\Eop$, however, 
is currently lacking in the literature.)

\begin{prop}\label{recursionmapping}
For the KN equation \eqref{KNeq}:
\newline
(i) if $P$ is any symmetry characteristic function 
and $Q$ is any adjoint-symmetry (or multiplier), 
the recursion operator \eqref{Rop1} generates 
respective hierarchies of symmetries and adjoint-symmetries given by 
$P_{(n)}= \Rop^nP$ and $Q_{(n)}= \Rop^*{}^nQ$, $n=0,1,2,\ldots$,
which have the relationship $Q_{(n)}= \Hop^{-1} P_{(n)}$. 
\newline
(ii) if $Q$ is any local multiplier, or if $P$ is any local Hamiltonian symmetry, 
then $Q_{(n)}$ and $P_{(n)}$, $n=0,1,2,\ldots$,
yield related hierarchies of local multipliers and local Hamiltonian symmetries.
The corresponding local conserved densities $T_{(n)}$ are given in terms of 
$Q_{(n)}= \Hop^{-1} P_{(n)}$ 
via the homotopy integral formula \eqref{TfromQ-homotopy} 
or the scaling formula \eqref{TfromQ-scaling}. 
\end{prop}

Here 
\begin{equation}\label{adRop1}
\Rop^* = D_x^4 - \tilde a_1D_x^3+\tilde a_2D_x^2-\tilde a_3D_x +\tilde a_4 +Q_{(1)}D_x^{-1}P_{(1)} +Q_{(2)}D_x^{-1}P_{(0)}  
\end{equation}
is the adjoint 4th order recursion operator of the KN equation \eqref{KNeq},
where the coefficients
\begin{equation}
\begin{aligned}
\tilde a_1 = a_1,
\quad
\tilde a_2 = a_2-3D_x a_1,
\quad
\tilde a_3 = a_3-2D_x a_2 +3D_x^2 a_1,
\quad
\tilde a_4 = a_4-D_x a_3 +D_x^2 a_2 -D_x^3 a_1
\end{aligned}
\end{equation}
are specific functions of $u$ and $x$-derivatives of $u$, 
which are given in terms of the coefficients $a_1,\ldots,a_4$ 
shown in Ref.~\cite{DemSok08}. 

Starting from the two translation symmetries \eqref{rootP0P1} as roots, 
the 4th order recursion operator \eqref{Rop1} of the KN equation 
generates intertwined hierarchies of local symmetries 
given by the characteristic functions
\begin{equation}
P_{(n,0)}=\Rop^n P_{(0)},
\quad
P_{(n,1)}=\Rop^n P_{(1)},
\quad
n=0,1,2,\ldots
\end{equation}
whose differential orders are, respectively, 
$1+4n$ and $3+4n$. 
At lowest order, 
$P_{(1,0)}=\Rop P_{(0)} = P_{(2)}$ 
has differential order 5, 
$P_{(1,1)}=\Rop P_{(1)} = P_{(3)}$ 
has differential order 7,
and so on,
all of which yield local symmetries. 
It is natural to organize them into a single hierarchy, 
denoted by 
\begin{equation}\label{localPhierarchy}
P_{(n)}=\begin{cases}
\Rop^{n/2} P_{(0)}, & n = \text{even}
\\
\Rop^{(n-1)/2} P_{(1)}, & n = \text{odd}
\end{cases}
\end{equation}
which has differential order $1+2n$, $n=0,1,2,\ldots$. 
The local Hamiltonian symmetries \eqref{P-1} and \eqref{P-2a} coincide with 
$\X_{(1)}=P_{(1)}\parder{u}$ and $\X_{(2)}=P_{(2)}\parder{u}$, respectively. 
All of the other local symmetries in this hierarchy 
also turn out to be Hamiltonian symmetries. 

By Proposition~\ref{recursionmapping}, 
the corresponding Hamiltonian conserved densities 
arise from the adjoint-symmetry counterpart of the hierarchy \eqref{localPhierarchy} 
as follows:
\begin{equation}\label{localQhierarchy}
E_u(H_{(n)}) =\Hop^{-1}P_{(n)} = Q_{(n)}
=\begin{cases}
\Rop^{*(n/2-1)} Q_{(2)}, & n = \text{even}
\\
\Rop^{*(n-1)/2} Q_{(1)}, & n = \text{odd}
\end{cases}
\end{equation}
which has differential order $2+2n$. 
At lowest order, 
$\Rop^* Q_{(1)} = Q_{(3)}$ 
has differential order 8, 
$\Rop^* Q_{(2)} = Q_{(4)}$ 
has differential order 10,
and so on,
all of which are local adjoint-symmetries that do not contain $t,x$. 
Correspondingly, 
$H_{(3)}$ has differential order 4, 
$H_{(4)}$ has differential order 5,
and so on. 
The scaling formula \eqref{TfromQ-scaling} can be used to obtain 
$H_{(n)}$ explicitly in terms of $Q_{(n)}$. 

Under the scaling equivalence transformation 
$t\rightarrow\lambda^3 t$, 
$x\rightarrow\lambda x$, 
$p(u)\rightarrow\lambda^{-4} p(u)$ 
of the KN equation \eqref{KNeq}, 
the scaling weight of each adjoint-symmetry in the hierarchy \eqref{localQhierarchy} is $\omega_{(n)}= -2n$, 
since $Q_{(1)}$ and $Q_{(2)}$ explicitly have the respective scaling weights
$\omega_{(1)}=-2$ and $\omega_{(2)}=-4$ 
while the scaling weight of the recursion operator $\Rop$ is $-4$. 
Since $u$ appears in these adjoint-symmetries only through $f(u)$, $f'(u)$, $f''(u)$, and so on, 
it is now convenient to express 
\begin{equation}
Q = \hat{Q}(v,v_x,v_{xx},\ldots,f,D_xf,D_x^2f,\ldots)
\end{equation}
by writing $v=u_x$, $v_x=u_{xx}$, and so on, 
and replacing $f'(u)=v^{-1}D_xf$, $f''(u)=v^{-2}D_x^2f-v_xv^{-3}D_xf$, 
and so on. 
The Hamiltonian multiplier relation $E_u(H) = Q$ 
combined with the variational identity \eqref{varid} 
then shows that every adjoint-symmetry will have the form 
\begin{equation}\label{Qhat-decompose}
\hat{Q} = v^{-1}\hat{Q}^fD_xf -D_x\hat{Q}^v
\end{equation}
where $\hat{Q}^f$ and $\hat{Q}^v$ are some functions of 
$v,f,v_x,D_xf,v_{xx},D_x^2f,\ldots$. 
As a result, 
when this decomposition is applied to each adjoint-symmetry $\hat{Q}_{(n)}$, 
the scaling formula \eqref{TfromQ-scaling} becomes 
$\hat{H}_{(n)} = (2n-1)^{-1}( 4v^{-1}\hat{Q}^f D_xf -D_x^{-1}(v\hat{Q}_{(n)}) )$, 
which yields the hierarchy of local conserved densities. 
It is interesting to note that the $D_x^{-1}$ term in this formula simplifies 
due to the relation 
$vD_x^{-1}(v\hat{Q}_{(n)}) =\Hop(v\hat{Q}_{(n)}) = \hat{P}_{(n)}$. 
Hence we obtain 
\begin{equation}\label{HfromQ-scaling}
H_{(n)} = (2n-1)^{-1}( 4f'(u)Q^f -P_{(n)}/u_x{} ), 
\quad
n=0,1,2,\ldots . 
\end{equation}

In particular, 
we find
\begin{equation}\label{H3}
\begin{aligned}
H_{(3)} = &
\frac{1}{2}\frac{u_{xxxx}^2}{u_x^2} +\frac{3}{2}\frac{u_{xxx}^3}{u_x^3} 
-\frac{19}{4}\frac{u_{xxx}^2u_{xx}^2}{u_x^4} 
+\frac{7}{6}\frac{u_{xxx}^2f(u)}{u_x^4} 
+\frac{45}{16}\frac{u_{xx}^6}{u_x^6} 
-\frac{259}{72}\frac{u_{xx}^4f(u)}{u_x^6} 
\\&\qquad
+\frac{35}{18}\frac{u_{xx}^3f'(u)}{u_x^4} 
+\frac{35}{36}\frac{u_{xx}^2f(u)^2}{u_x^6} 
-\frac{7}{9}\frac{u_{xx}^2f''(u)}{u_x^2}
+\frac{1}{54}\frac{f(u)^3}{u_x^6}
\\&\qquad
-\frac{7}{54}\frac{f'(u)^2+2f(u)f''(u)}{u_x^2}
-\frac{7}{9}f''''(u)u_x^2 
\end{aligned}
\end{equation}
which was first obtained in Ref.~\cite{DemSok08}, 
and 
\begin{equation}\label{H4}
\begin{aligned}
H_{(4)} = &
\frac{1}{2}\frac{u_{xxxxx}^2}{u_x^2} 
+8\frac{u_{xxxx}^2u_{xxx}}{u_x^3} 
-\frac{33}{4}\frac{u_{xxxx}^2u_{xx}^2}{u_x^4} 
+\frac{3}{2}\frac{u_{xxxx}^2f(u)}{u_x^4} 
+\frac{85}{8}\frac{u_{xxx}^4}{u_x^4} 
-\frac{323}{4}\frac{u_{xxx}^3u_{xx}^2}{u_x^5} 
\\&\qquad
+\frac{47}{6}\frac{u_{xxx}^3f(u)}{u_x^5} 
+\frac{2103}{16}\frac{u_{xxx}^2u_{xx}^4}{u_x^6} 
-\frac{201}{4}\frac{u_{xxx}^2u_{xx}^2f(u)}{u_x^6} 
+15\frac{u_{xxx}^2u_{xx}f'(u)}{u_x^4} 
\\&\qquad
+\frac{7}{4}\frac{u_{xxx}^2f(u)^2}{u_x^6} 
-3\frac{u_{xxx}^2f''(u)}{u_x^2} 
-\frac{7875}{128}\frac{u_{xx}^8}{u_x^8} 
+\frac{3229}{48}\frac{u_{xx}^6f(u)}{u_x^8} 
-\frac{287}{8}\frac{u_{xx}^5f'(u)}{u_x^6} 
\\&\qquad
-\frac{1645}{144}\frac{u_{xx}^4f(u)^2}{u_x^8} 
+\frac{32}{3}\frac{u_{xx}^4f''(u)}{u_x^4} 
+\frac{175}{18}\frac{u_{xx}^3f(u)f'(u)}{u_x^6} 
-\frac{20}{9}\frac{u_{xx}^3f'''(u)}{u_x^2} 
\\&\qquad
+\frac{35}{36}\frac{u_{xx}^2f(u)^3}{u_x^8} 
-\frac{1}{12}\frac{u_{xx}^2(23f'(u)^2+44f(u)f''(u))}{u_x^2}
-\frac{7}{6}f''''(u)u_{xx}^2 
+\frac{5}{648}\frac{f(u)^4}{u_x^8}
\\&\qquad
-\frac{1}{6}\frac{f(u)f'(u)^2+f(u)^2f''(u)}{u_x^4}
+\frac{278}{189}f(u)f''''(u)+\frac{415}{189}f'(u)f'''(u)
+\frac{265}{189}f''(u)^2 
\end{aligned}
\end{equation}
which has not appeared previously in the literature.

\subsection{Hierarchy of nonlocal symmetries}

None of the local Hamiltonian symmetries 
given by the hierarchy \eqref{localPhierarchy}
contain $t,x$. 
In contrast, 
the Hamiltonian symmetry \eqref{P-3a} that contains $t,x$ 
is a new nonlocal symmetry. 
By part (i) of Proposition~\ref{recursionmapping}, 
the recursion operator \eqref{Rop1} of the KN equation 
can be used to generate a hierarchy of similar nonlocal symmetries.

\begin{thm}\label{nonlocalsymms}
The KN equation \eqref{KNeq} admits a hierarchy of 
nonlocal symmetries
\begin{equation}\label{nonlocalPhierarchy}
\hat\Y_{(n)}= \Rop^n\big( 3t\Rop u_x +x u_t + u_{xx} - u_xD_x^{-1} H_{(1)} \big)\parder{u},
\quad
n=0,1,2,\ldots
\end{equation}
where the characteristic function can be expressed as
\begin{equation}\label{nonlocalP}
\wtil P_{(n)}= -3tP_{(2n+2)} +xP_{(2n+1)} +\text{ terms that do not contain $t,x$} 
\end{equation}
in terms of the characteristic functions \eqref{localPhierarchy}
for the hierarchy of local symmetries $\X_{(n)}=P_{(n)}\parder{u}$, 
$n=0,1,2,\ldots$.
\end{thm} 

In particular, the root symmetry has the characteristic function
\begin{equation}\label{nonlocalP0}
\wtil P_{(0)} = -3tP_{(2)} + xP_{(1)} -P_{(0)}D_x^{-1}H_{(1)}+u_{xx}
\end{equation}
and the next symmetry is given by 
\begin{equation}\label{nonlocalP1}
\begin{aligned}
\wtil P_{(1)} = & 
-3tP_{(4)} + xP_{(3)} 
+P_{(2)}D_x^{-1}H_{(1)}+3P_{(1)}D_x^{-1}H_{(2)}+5P_{(0)}D_x^{-1}H_{(3)}
\\&\qquad
+5u_{xxxxxx}
-29\frac{u_{xxxxx}u_{xx}}{u_x} 
-47\frac{u_{xxxx}u_{xxx}}{u_x} 
+ 103\frac{u_{xxxx}u_{xx}^2}{u_x^2} 
- \frac{28}{3}\frac{u_{xxxx}f(u)}{u_x^2} 
\\&\qquad
+\frac{291}{2}\frac{u_{xxx}^2 u_{xx}}{u_x^2} 
-277\frac{u_{xxx} u_{xx}^3}{u_x^3} 
+ \frac{200}{3}\frac{u_{xxx} u_{xx}f(u)}{u_x^3} 
-13\frac{u_{xxx} f'(u)}{u_x} 
+ 99\frac{u_{xx}^5}{u_x^4} 
\\&\qquad
-\frac{664}{9}\frac{u_{xx}^3f(u)}{u_x^4} 
+\frac{59}{2}\frac{u_{xx}^2f'(u)}{u_x^2} 
+\frac{16}{3}\frac{u_{xx}f(u)^2}{u_x^4} 
-\frac{52}{27}\frac{f(u)f'(u)}{u_x^2} 
-\frac{65}{9}u_{xx}f''(u) . 
\end{aligned}
\end{equation}

An interesting question is 
to determine if these nonlocal symmetries are Hamiltonian, 
since a generalization of part (ii) of Proposition~\ref{recursionmapping}
to the nonlocal case does not always hold. 
This requires determining if, for $n=1,2,\ldots$, 
$\wtil P_{(n)} = \Hop(\delta\wtil T_{(n)}/\delta u)$ 
holds for some nonlocal Hamiltonians $\wtil T_{(n)}$. 
Note, from expressions \eqref{P-3a}, the root symmetry \eqref{nonlocalP0} itself
is Hamiltonian, as given by 
$\wtil P_{(0)} = \Hop \wtil Q_{(0)}$
with 
\begin{equation}\label{nonlocalQ0}
\wtil Q_{(0)} 
= -3t Q_{(2)}+x Q_{(1)} 
+ 2\frac{u_{xxx}}{u_x^2}  -3\frac{u_{xx}^2}{u_x^3} +\frac{2}{3}\frac{f(u)}{u_x^3} 
= \delta\wtil T_{(0)}/\delta u
\end{equation}
where $\wtil T_{(0)} = -3t H_{(2)}+x H_{(1)}$. 
Then, part (i) of Proposition~\ref{recursionmapping} shows that 
\begin{equation}\label{nonlocalQhierarchy}
\wtil P_{(n)} = \Hop \wtil Q_{(n)},
\quad
\wtil Q_{(n)} = \Rop^*{}^n \wtil Q_{(0)}, 
\quad
n=1,2,\ldots
\end{equation}
where each $\wtil Q_{(n)}$ is an adjoint-symmetry (or cosymmetry) of the KN equation \eqref{KNeq}. 
Consequently, 
a Hamiltonian exists iff these adjoint-symmetries have the form 
$\wtil Q_{(n)} = \delta\wtil T_{(n)}/\delta u$ 
for some nonlocal functions $\wtil T_{(n)}$. 
In general, 
it seems natural to expect that the nonlocality in $\wtil T_{(n)}$ 
should involve the same nonlocal expressions appearing in $\wtil Q_{(n)}$. 

To begin, 
we consider the first adjoint-symmetry in the hierarchy \eqref{nonlocalQhierarchy},
which is given by 
\begin{equation}\label{nonlocalQ1}
\begin{aligned}
\wtil Q_{(1)}
& = -3t Q_{(4)}+x Q_{(3)} +Q_{(2)}D_x^{-1}H_{(1)}+3Q_{(1)}D_x^{-1}H_{(2)}
\\&\qquad
+6\frac{u_{xxxxxxx}}{u_x^2}
-41\frac{u_{xxxxxx}u_{xx}}{u_x^3} 
-90\frac{u_{xxxxx}u_{xxx}}{u_x^3} 
+192\frac{u_{xxxxx}u_{xx}^2}{u_x^4} 
-12\frac{u_{xxxxx}f(u)}{u_x^4} 
\\&\qquad
-60\frac{u_{xxxx}^2}{u_x^3} 
+710\frac{u_{xxxx}u_{xxx}u_{xx}}{u_x^4}
-699\frac{u_{xxxx}u_{xx}^3}{u_x^5} 
+\frac{352}{3}\frac{u_{xxxx}u_{xx}f(u)}{u_x^5} 
-27\frac{u_{xxxx}f'(u)}{u_x^3} 
\\&\qquad
+164\frac{u_{xxx}^3}{u_x^4} 
-1488\frac{u_{xxx}^2 u_{xx}^2}{u_x^5} 
+76\frac{u_{xxx}^2f(u)}{u_x^5} 
+1908\frac{u_{xxx} u_{xx}^4}{u_x^6} 
-588\frac{u_{xxx} u_{xx}^2f(u)}{u_x^6} 
\\&\qquad
+182\frac{u_{xxx} u_{xx}f'(u)}{u_x^4} 
+8\frac{u_{xxx} f(u)^2}{u_x^6} 
-\frac{232}{9}\frac{u_{xxx} f''(u)}{u_x^2} 
-603\frac{u_{xx}^6}{u_x^7} 
+\frac{1372}{3}\frac{u_{xx}^4f(u)}{u_x^7} 
\\&\qquad
-205\frac{u_{xx}^3f'(u)}{u_x^5} 
-\frac{112}{3}\frac{u_{xx}^2f(u)^2}{u_x^7} 
+51\frac{u_{xx}^2f''(u)}{u_x^3} 
+\frac{188}{9}\frac{u_{xx}f(u)f'(u)}{u_x^5} 
-\frac{79}{9}\frac{u_{xx}f'''(u)}{u_x} 
\\&\qquad
+\frac{8}{27}\frac{f(u)^3}{u_x^7} 
-\frac{88}{27}\frac{f(u)f''(u)}{u_x^3} 
-\frac{5}{3}\frac{f'(u)^2}{u_x^3} 
+\frac{14}{3}u_{x}f''''(u) . 
\end{aligned}
\end{equation}
The nonlocal terms in this adjoint-symmetry involve only 
$D_x^{-1}H_{(1)}$ and $D_x^{-1}H_{(2)}$.
We now formulate Helmholtz conditions that are necessary and sufficient
for the variational property 
$\wtil Q_{(1)} = \delta\wtil T_{(1)}/\delta u$ to hold, 
assuming that all nonlocal dependence in $\wtil T_{(1)}$ 
involves only the expressions $D_x^{-1}H_{(1)}$ and $D_x^{-1}H_{(2)}$.
The conditions are simply
\begin{equation}
\wtil Q_{(1)}'(w) = \wtil Q_{(1)}^{\prime *}(w)
\end{equation}
where $w$ is an arbitrary function of $t,x$. 
Here the prime denotes the Frechet derivative with respect to $u$,
and the star denotes the formal adjoint (as defined by integration by parts),
where $(BD_x^{-1}A)'(w) = BD_x^{-1}(A'(w)) + (D_x^{-1}A)B'(w)$
and $(BD_x^{-1}A)^{\prime *}(w) = B^{\prime *}(wD_x^{-1}A) - A^{\prime *}(D_x^{-1}(wB))$
for any local differential functions $A,B$. 
These expressions can be simplified by using the variational identities
$A'(w)= wE_u(A) + D_x(wE_u^{(1)}(A)) + D_x^2(wE_u^{(2)}(A)) +\cdots$
and 
$A^{\prime *}(w) = wE_u(A) -E_u^{(1)}(A)D_xw + E_u^{(2)}(A)D_x^2w +\cdots$,
which yields 
\begin{align}
&\begin{aligned}
(BD_x^{-1}A)'(w) = & 
(D_x^{-1}A)B'(w) + BD_x^{-1}(wE_u(A)) 
\\&\qquad
+ BwE_u^{(1)}(A)) + BD_x(wE_u^{(2)}(A)) +\cdots , 
\end{aligned}
\\
&\begin{aligned}
(BD_x^{-1}A)^{\prime *}(w) = & 
(D_x^{-1}A) B^{\prime *}(w) -(D_x^{-1}B)E_u(A) 
+ wBE_u^{(1)}(A) -AE_u^{(1)}(wB) 
\\&\qquad
- D_x(wB)E_u^{(2)}(A) +D_xA(E_u^{(2)}(wB) +\cdots .
\end{aligned}
\end{align}
From expression \eqref{nonlocalQ1}, we then find that 
\begin{equation}
\begin{aligned}
\wtil Q_{(1)}'(w) - \wtil Q_{(1)}^{\prime *}(w) & = 
2Q_{(1)}D_x^{-1}(wQ_{(2)}) -2Q_{(2)}D_x^{-1}(wQ_{(1)}) 
\\&\qquad
+4\frac{1}{u_x^2}D_x^7 w -28\frac{u_{xx}}{u_x^3}D_x^6 w +\cdots 
\neq 0 . 
\end{aligned}
\end{equation}
Hence, the Helmholtz conditions fail to hold, 
which implies that $\wtil Q_{(1)}$ does not have the form 
$\delta\wtil T_{(1)}/\delta u$ for any function $\wtil T_{(1)}$ 
whose nonlocal dependence involves only the expressions $D_x^{-1}H_{(1)}$ and $D_x^{-1}H_{(2)}$.

As a check, 
we next look for $\wtil T_{(1)}$ by adapting the scaling formula \eqref{HfromQ-scaling}, 
without the assumption that the only nonlocality in $\wtil T_{(1)}$ 
should involve $D_x^{-1}H_{(1)}$ and $D_x^{-1}H_{(2)}$.

It is easy to see $\wtil Q_{(0)}$ has scaling weight $\wtil\omega_{(0)}=-1$.
Hence, the scaling weight of $\wtil Q_{(n)}$ for $n=1,2,\ldots$ 
is $\wtil\omega_{(n)}=-1-4n\neq 0$,
since $\Rop^*$ has the scaling weight $-4$. 
Then the scaling formula \eqref{HfromQ-scaling} becomes
\begin{equation}\label{HfromnonlocalQ-scaling}
\wtil T_{(n)} = (4n)^{-1}( 4f'(u)\widehat{\wtil Q}{}^f_{(n)} -D_x^{-1}(v\widehat{\wtil Q}{}_{(n)}) ),
\quad
n=1,2,\ldots 
\end{equation}
where $\widehat{\wtil Q}{}^f_{(n)}$ and $\widehat{\wtil Q}{}_{(n)}$ 
are given by the decomposition \eqref{Qhat-decompose}. 
We apply this formula for $n=1$, 
after decomposing $\widehat{\wtil Q}{}_{(1)}$ to get  
\begin{equation}\label{Q1f}
\begin{aligned}
\widehat{\wtil Q}{}_{(1)}^f
& = -3t Q_{(4)}^f+x Q_{(3)}^f +Q_{(2)}^f D_x^{-1}H_{(1)}+3Q_{(1)}^f D_x^{-1}H_{(2)}
\\&\qquad
-\frac{7}{3}\frac{v_{xxx}}{v^3} 
+\frac{79}{9}\frac{v_{xx}v_x}{v^4} 
-\frac{20}{3}\frac{v_x^3}{v^5} 
+2\frac{v_x f}{v^5} 
-\frac{2}{9}\frac{D_x f}{v^4} . 
\end{aligned}
\end{equation} 
A direct evaluation of the terms in the scaling formula shows 
\begin{equation}
\begin{aligned}
4f'(u)\widehat{\wtil Q}{}^f_{(1)} -D_x^{-1}(v\widehat{\wtil Q}{}_{(1)}) = & 
D_x\bigg( \text{ local terms } 
+ 5xD_x^{-1}H_3 + 3(D_x^{-1}H_1)D_x^{-1}H_2 -3\frac{v_x}{v}D_x^{-1}H_2
\\&\qquad
+\Big( \frac{20}{9}\frac{D_xf}{v^2}  + \frac{5}{9}\frac{v_xf}{v^2} 
+\frac{3}{2}\frac{v_x^3}{v^3} -4\frac{v_xv_{xx}}{v^2} +\frac{v_{xxx}}{v}\Big)D_x^{-1}H_1 \bigg)
\end{aligned}
\end{equation} 
is a total $x$-derivative. 
This implies that $\wtil Q_{(1)} = \delta\wtil T_{(1)}/\delta u$ 
does not hold for any nonlocal function $\wtil T_{(1)}$. 

Therefore, surprisingly,  
these results show that $\wtil P_{(1)}$ is not Hamiltonian. 
The same conclusion can be expected to hold for the entire hierarchy \eqref{nonlocalQhierarchy}.

\section{Concluding remarks}
\label{remarks}

The main importance of the hierarchy of new nonlocal symmetries \eqref{nonlocalPhierarchy}
found in this paper for the KN equation 
is that the root symmetry \eqref{nonlocalP0} can be shown
to play the role of a master symmetry \cite{Olv} 
which generates the entire hierarchy of local Hamiltonian symmetries \eqref{localPhierarchy} of the KN equation. 
A full discussion of this result will be presented in a subsequent paper \cite{mastersymm}. 

Our derivation of the nonlocal symmetries \eqref{nonlocalPhierarchy} 
utilizes the basic Hamiltonian structure of the KN equation 
together with its 4th order recursion operator \eqref{Rop1}. 
The KN equation admits another recursion operator, 
which is 6th order and is related to the 4th order one by an elliptic curve \cite{DemSok08}. 
This other recursion operator can also be used to generate 
a hierarchy of nonlocal symmetries which will intertwine 
with the hierarchy \eqref{nonlocalPhierarchy}.
To understand this intertwining,
and to look possibly for more nonlocal symmetries, 
it would be interesting to extend the methods of the present paper 
to carry out a systematic computational classification of 
nonlocal symmetries and nonlocal conservation laws for the KN equation.

\section*{Acknowledgements}

S.C. Anco and T. Wolf are each supported by an NSERC research grant. 
E.D. Avdonina, A. Gainetdinova, L.R. Galiakberova, N.H. Ibragimov
acknowledge the financial support of the Government of Russian Federation 
through Resolution No.\ 220, Agreement No.\ 11.G34.31.0042.
The referees are thanked for helpful remarks which have improved this paper.

 \end{document}